\documentclass[aps,onecolumn,nofootinbib,tightenlines,superscriptaddress]{revtex4}

\usepackage{amssymb,amstext,amsmath,amsthm,slashed}
\usepackage[dvips]{graphicx}
\usepackage{latexsym}
\usepackage{psfrag}
\usepackage{amsfonts}
\usepackage{color}
\usepackage{verbatim}
\usepackage{bbm}
\usepackage{dcolumn}
\usepackage{tabularx}
\usepackage{leftidx}
\usepackage{tensor}
\usepackage{boxedminipage}
\usepackage{fancyvrb}
\usepackage{float}
\usepackage{wrapfig}
\usepackage{booktabs}
\usepackage{multirow}
\usepackage{subfigure}
\usepackage{dsfont} 
\usepackage{mathrsfs}
\usepackage{scalerel,stackengine}
\usepackage{amsthm}
\usepackage[dvipsnames]{xcolor}
\usepackage[hidelinks]{hyperref}

\theoremstyle{plain}

\newtheorem{Def}{Definition}[section]

\newtheorem*{theorem*}{Theorem}

\newtheorem{Remark}[Def]{Remark}

\newcommand{\1}{\mbox{\rm 1 \hspace{-1.05 em} 1}}

\newcommand*{\medcup}{\mathbin{\scalebox{1.5}{\ensuremath{\cup}}}}

\newcommand\stoptoc{\let\addcontentsline\nocontentsline}
\newcommand\resumetoc{\let\addcontentsline\origcontentsline}
\newcommand{\nocontentsline}[3]{}
\let\origcontentsline\addcontentsline
\begin{document}

\title{The Construction and Application of Penrose Diagrams, with a Focus \\ on the Maximally Analytically Extended Schwarzschild Spacetime\vspace{0.1cm}}


\author{Christian R\"oken\vspace{0.25cm}} 

\email{croeken@uni-bonn.de}

\affiliation{Argelander Institute for Astronomy - University of Bonn, 53121 Bonn, Germany \vspace{0.25cm}}

\affiliation{Lichtenberg Group for History and Philosophy of Physics - Faculty of Philosophy, University of Bonn, 53113 Bonn, Germany \vspace{0.25cm}}


\date{July 2025 / February 2026}

\begin{abstract}
\vspace{0.4cm} \noindent \textbf{\footnotesize ABSTRACT.} \, We present a detailed, mathematically rigorous description of the construction procedure of Penrose diagrams for the example of the maximal analytic extension of the exterior Schwarzschild spacetime. To this end, we first outline the central idea underlying Penrose diagrams, state the general requirements on the spacetimes to be visualized, and give a definition of Penrose diagrams. We then construct the Penrose diagram of the maximally analytically extended Schwarzschild spacetime and discuss its components and characteristics. As an application, we work out the differences between the Eddington--Finkelstein and Penrose coordinate representations of the Schwarzschild spacetime by visually analyzing---and comparing---the Penrose diagram of the maximally analytically extended Schwarzschild spacetime equipped with, on the one hand, a foliation by the level sets of the Eddington--Finkelstein time coordinate and, on the other hand, a foliation by the level sets of the Penrose null coordinate. Through the whole of the paper, we provide explanatory accounts of the relevant parts of the seminal research papers on the exterior Schwarzschild solution pertaining to coordinatizations and possible extensions by Schwarzschild himself, Kruskal, Eddington, Finkelstein, and Penrose. This paper is primarily of pedagogical nature aimed at graduate students in physics and applied mathematics, serving mainly as an introduction to Penrose diagrams along with descriptions and analyses of coordinate representations and extensions of the exterior Schwarzschild spacetime.
\end{abstract}

\makeatletter
\def\l@subsubsection#1#2{}
\makeatother

\vspace{0.1cm}

\maketitle

\tableofcontents

\section{Introduction} \label{SectionI}

\noindent In the framework of general relativity, and more generally in the field of differential geometry, many questions of interest are concerned with the global causal structure and the infinities of $4$-dimensional curved spacetimes, such as the question of the geodesic completeness of spacetimes with low differentiability or of the geometric and topological spacetime requirements for particular causality conditions to be satisfied. A particularly useful visualization tool in this regard are Penrose diagrams, which are faithful $2$-dimensional, finite-sized diagrammatic representations that contain determining information on the global causal structure and on possible infinities of entire (or particular subsets of) $4$-dimensional spacetimes. The general idea underlying Penrose diagrams was first introduced by Roger Penrose in 1963/64 in \cite{Penrose0, Penrose1} and developed further into its present-day form by Carter and Walker in \cite{Carter} and \cite{Walker}, respectively. Penrose diagrams may be considered as the natural finite-sized Lorentzian generalization of the usual Minkowski diagrams of special relativity or as a finite-sized variant of the Kruskal diagrams of general relativity, where infinity is now mapped onto the finite boundary of the diagram, allowing for proper studies of asymptotic properties of spacetimes. These diagrams can, a priori, be constructed for all kinds of spacetimes, ranging from simple vacuum spacetimes to spacetimes generated by nontrivial, dynamically evolving energy-momentum distributions. But for them to be meaningful, they have to contain appropriate information on the full $4$-dimensional spacetimes so that useful inferences about the underlying geometries can be drawn. And although Penrose diagrams are by themselves not capable of directly visualizing the energy-momentum distribution inherent to any nonvacuum spacetime for they are not based on more than restrictions or projections, a certain coordinatization, and a compactification of the underlying spacetimes,\footnote{Energy-momentum distributions and their possible spatio-temporal evolutions are usually incorporated into Penrose diagrams manually afterwards.} they can still reveal the energy-momentum distributions' influences on the causal structures of spacetimes in qualitative analyses.

The aim of this paper is to give a comprehensive pedagogical but at the same time mathematically rigorous account of Penrose diagrams for graduate students in physics and applied mathematics having solid backgrounds in general relativity and differential geometry, with an explicit construction of such a diagram carried out for the example of the maximally analytically extended Schwarzschild spacetime.\footnote{For basic introductions to general relativity and differential geometry see the standard textbooks \cite{HawkingEllis, Wald, MTW} and \cite{ONeill2, ONeill3, Lee}, respectively.} Therefore, we also focus on the motivations, definitions, and interpretations of the various coordinate representations and related extensions of the exterior Schwarzschild spacetime relevant for the construction of coordinates that not only cover its maximal analytic extension but also form a proper basis for the associated Penrose diagram. As an application for Penrose diagrams, we address how they can be used to illustrate, and thus identify, the differences between two particular coordinate representations of the Schwarzschild spacetime. In more detail, in the preliminary section \ref{SubsectionIIA}, we first recall the details and main results of Schwarzschild's acclaimed 1916 paper \cite{Schwarzschild1} on the derivation of the external gravitational field of a massive point particle within the theory of general relativity before giving a modern account of this so-called exterior Schwarzschild solution and the construction of the corresponding maximal analytic extension, for which we recapitulate Kruskal's 1960 paper \cite{Kruskal}, in Section \ref{SubsectionIIB}. In Section \ref{SectionIII}, we discuss the general idea behind Penrose diagrams, formulate conditions on the spacetimes to be represented, examine the characteristics of Penrose diagrams, and then give a rigorous definition of these diagrams. Subsequently, in Section \ref{SectionIV}, we get to a detailed conceptual step-by-step description of the construction procedure of Penrose diagrams, focusing on the example of the maximally analytically extended Schwarzschild spacetime and the required coordinate representations. For an application of this Penrose diagram, we discuss in Section \ref{SubsectionVA}, on the one hand, the relevant parts pertaining to the canonical Eddington--Finkelstein coordinate representation of the exterior Schwarzschild solution introduced in Eddington's 1924 paper \cite{Eddington} and Finkelstein's 1958 paper \cite{Finkelstein}, and, on the other hand, the details of the related Penrose coordinate representation employed in Penrose's 1969 paper \cite{Penrose3}.\footnote{Here, we shall refer to the time-reversed null variant of the Eddington--Finkelstein coordinates, although similarly coined ``Eddington--Finkelstein coordinates'' in \cite{Penrose3} by Penrose himself, as ``Penrose coordinates,'' for they are sufficiently different in comparison to the coordinates used by Eddington and Finkelstein (for details see Section \ref{SubsectionVA}).} With this at hand, we finally answer the question of how precisely these two coordinate representations differ by visually analyzing---and then comparing the characteristics of---the families of level sets of the respective Eddington--Finkelstein time and Penrose null coordinates foliating the combined exterior and interior black hole regions of the Penrose diagram of the maximal analytic extension of the Schwarzschild spacetime in Section \ref{SubsectionVB}.

\section{Preliminaries} \label{SectionII}

\subsection{Historical Remarks on the Exterior Schwarzschild Solution} \label{SubsectionIIA}

\noindent In order to present the construction procedure of Penrose diagrams on the basis of the example of the maximally analytically extended Schwarzschild spacetime in a comprehensible manner, it is advantageous to have a basic understanding of the nature and the properties of this particular spacetime. Thus, we first recall the relevant background of Schwarzschild's original solution. In 1916, Karl Schwarzschild published two very influential papers, viz.\ \cite{Schwarzschild1} and \cite{Schwarzschild2}, on the gravitational field of a spherically symmetric body within the framework of Einstein's then newly established theory of general relativity. To be more precise, the aim of the first of these papers, which is pertinent to the present study, was the derivation of an explicit analytical solution of the vacuum Einstein field equations that accounts for the gravitational field generated by an isolated, spherically symmetric point particle of invariant mass, located at the origin of a Cartesian coordinate system $(x^i)_{i \in \{1, 2, 3\}} \in \mathbb{R}^3 \backslash \{\boldsymbol{0}\}$ for times $x^4 \in \mathbb{R}$ together covering a connected $4$-manifold $\mathfrak{M}$, for the study of the motion of a test particle in such a gravitational field.\footnote{For a historical overview of this paper see, e.g., \cite{Eisenstaedt1, Eisenstaedt2}. A comprehensive physical review can be found in \cite{HeinickeHehl}.}$^{,}$\footnote{The origin at $(x^1, x^2, x^3) = \boldsymbol{0}$, which is, according to Schwarzschild, the location of the gravitational-field-generating point particle (cf.\ Remark \ref{RemLocPMS}), is not contained in the range of the Cartesian coordinates, as for Schwarzschild only the gravitational field outside the point particle is of interest.} To this end, Schwarzschild was seeking a metric $\boldsymbol{g}$ that at every event satisfies the unimodularity condition $|\textnormal{det}(\boldsymbol{g})| = - 1$ and the field equations 
\begin{equation} \label{UnimodularEFE}
\partial_{\alpha} \Gamma^{\alpha}_{\mu \nu} + \Gamma^{\alpha}_{\mu \beta} \Gamma^{\beta}_{\nu \alpha} = 0 \, , \quad \alpha, \beta, \mu, \nu \in \{1, 2, 3, 4\} \, , 
\end{equation}
where   
\begin{equation*}
\Gamma^{\alpha}_{\mu \nu} = - \frac{1}{2} \, g^{\alpha \beta} (\partial_{\nu} g_{\mu \beta} + \partial_{\mu} g_{\nu \beta} - \partial_{\beta} g_{\mu \nu})
\end{equation*}
are the usual Christoffel symbols of the second kind,\footnote{Schwarzschild, in fact, employed the earlier unimodular vacuum Einstein field equations introduced in \cite{Einstein1} instead of the fully covariant vacuum Einstein field equations defined in \cite{Einstein2}. This, however, does not pose a problem for his later analysis of the gravitational field of a point particle, as the resulting solution indeed satisfies both of these sets of equations. Schwarzschild's paper was followed shortly after by Droste's work \cite{Droste} that more rigorously deals with the same problem within the generally covariant framework.}$^{,}$\footnote{Schwarzschild himself did not specify any details of the underlying manifold $\mathfrak{M}$. Moreover, he gave no mathematical definition of the metric $\boldsymbol{g}$ except for stating that it is ``a function of the variables $\boldsymbol{x}$,'' that is, of the spacetime coordinates. However, from a modern differential geometric point of view, the manifold may be considered as being smooth, and the metric may be interpreted as a section of the tensor product bundle of the cotangent bundle $T^{\star}\mathfrak{M}$ with itself, so that Schwarzschild was ultimately looking for solutions $\bigl\{\boldsymbol{g} \in \Gamma(\mathfrak{M}, T^{\star}\mathfrak{M} \otimes T^{\star}\mathfrak{M}) \, \big| \, |\textnormal{det}(\boldsymbol{g})| = - 1\bigr\}$ of the unimodular Einstein field equations (\ref{UnimodularEFE}) fulfilling the conditions ($\mathcal{C}1$)--($\mathcal{C}4$) stated below.} while at the same time it is being subject to the following conditions: 
\begin{itemize}
\item[($\mathcal{C}1$)] All metric components $(g_{\mu \nu})_{\mu, \nu \in \{1, 2, 3, 4\}}$ are independent of the time coordinate $x^4$.
\item[($\mathcal{C}2$)] The spatio-temporal metric components $(g_{i, 4})_{i \in \{1, 2, 3\}}$ are identically zero. 
\item[($\mathcal{C}3$)] The metric is spatially symmetric around the origin in the sense that it is invariant under special orthogonal transformations $T \in \textnormal{SO}(3)$ [i.e., rotations] applied to the coordinates $(x^i)_{i \in \{1, 2, 3\}}$. 
\item[($\mathcal{C}4$)] Except for the four diagonal metric components $(g_{\mu \mu})_{\mu \in \{1, 2, 3, 4\}}$, which are assumed to tend to the constants $g_{1 1} = g_{2 2} = g_{3 3} = - g_{4 4} = - 1$, all metric components vanish at infinity. 
\end{itemize}
Here, Conditions ($\mathcal{C}1$) and ($\mathcal{C}2$) imply that the spacetime build on the metric $\boldsymbol{g}$ is time-independent and irrotational, respectively,\footnote{By assuming that the coordinate $x^4$ is \textit{globally} timelike, such a spacetime may be referred to as \textit{static} (in the modern sense). This can be more properly, i.e., in a coordinate-independent way, inferred from the fact that in this case the vector field $\partial_{x^4}$ is a globally nonvanishing, timelike and irrotational Killing field.} whereas Conditions ($\mathcal{C}3$) and ($\mathcal{C}4$) together entail that it is spherically symmetric and asymptotically flat.\footnote{Mathematically more precise, spherical symmetry can be defined as the special orthogonal group $\textnormal{SO}(3)$ acting on $\mathfrak{M}$ as the group of isometries $\mathfrak{I} \colon \mathfrak{M} \times \textnormal{SO}(3) \rightarrow \mathfrak{M}$, where the set of fixed points $\mathfrak{I}\bigl((x^{\mu})_{\mu \in \{1, 2, 3, 4\}}, T\bigr) = (x^{\mu})_{\mu \in \{1, 2, 3, 4\}}$ for all $T \in \textnormal{SO}(3)$ is a timelike line referred to as the \textit{axis of symmetry}. Asymptotic flatness simply means that at spacelike infinity the metric becomes Minkowskian, i.e., $\lim_{|\boldsymbol{x}| \rightarrow \infty} \boldsymbol{g} = \boldsymbol{\eta} = \textnormal{d}x^4 \otimes \textnormal{d}x^4 - \textnormal{d}x^1 \otimes \textnormal{d}x^1 - \textnormal{d}x^2 \otimes \textnormal{d}x^2 - \textnormal{d}x^3 \otimes \textnormal{d}x^3$.} Employing the notation $(x = x^1, y = x^2, z = x^3)$ and $t = x^4$, and using the radial distance $r = \sqrt{x^2 + y^2 + z^2}$, Schwarzschild begins his derivation by making the metric ansatz
\begin{equation} \label{MetricAnsatz}
\boldsymbol{g} = F(r) \, \textnormal{d}t \otimes \textnormal{d}t - G(r) \, [\textnormal{d}x \otimes \textnormal{d}x + \textnormal{d}y \otimes \textnormal{d}y + \textnormal{d}z \otimes \textnormal{d}z] - H(r) \, [x \textnormal{d}x + y \textnormal{d}y + z \textnormal{d}z] \otimes [x \textnormal{d}x + y \textnormal{d}y + z \textnormal{d}z] \, ,
\end{equation}
which satisfies all of the above conditions assuming that at spatial infinity $F, G \rightarrow 1$ and $H \rightarrow 0$. In a next step, possibly for simplicity, he applies the transformation from Cartesian coordinates to spherical polar coordinates 
\begin{equation} \label{T1S}
\mathfrak{T}^{(1)} \colon
\begin{cases}
\, \mathbb{R} \backslash \{0\} \times \mathbb{R} \backslash \{0\} \times \mathbb{R} \backslash \{0\} \rightarrow \mathbb{R}_{> 0} \times (0, \pi) \times [0, 2 \pi)  \\[0.25cm]
\hspace{2.65cm} (x, y, z) \mapsto (r, \vartheta, \phi) 
\end{cases} 
\end{equation}
with
\begin{equation*}
\vartheta = \arccos{\biggl(\frac{z}{\sqrt{x^2 + y^2 + z^2}}\biggr)} \quad \textnormal{and} \quad \phi = \textnormal{sgn}(y) \biggl[\arccos{\biggl(\frac{x}{\sqrt{x^2 + y^2}}\biggr)} - \pi\biggr] + \pi \, .
\end{equation*}
This transformation, however, gives rise to coordinates that are not unimodular as required by the field equations defined in Equation (\ref{UnimodularEFE}) [cf.\ Footnote 6]. To resolve this problem, Schwarzschild rewrites the spatial volume element represented in the spherical polar coordinates (\ref{T1S}) in the form 
\begin{equation*}
\textnormal{vol}_3 = r^2 \sin{(\vartheta)} \, \textnormal{d}r \hspace{0.03cm} \textnormal{d}\vartheta \hspace{0.03cm} \textnormal{d}\phi = \textnormal{d}\biggl(\frac{r^3}{3}\biggr) \textnormal{d}\bigl(- \cos{(\vartheta)}\bigr) \textnormal{d}\phi = \textnormal{d}x_1 \textnormal{d}x_2 \textnormal{d}x_3 \, ,
\end{equation*}
which makes it possible to directly read out the transformation from spherical polar coordinates to unimodular spherical polar coordinates\footnote{This approach is viable because of the particular product structure of the Jacobian determinant for the change from Cartesian to spherical polar coordinates.}$^{,}$\footnote{Schwarzschild's unimodular spherical polar coordinates $(x_1, x_2, x_3)$ must not be confused with the usual covariant form of his initially chosen Cartesian coordinates $(x^1, x^2, x^3)$.}
\begin{equation*} 
\mathfrak{T}^{(2)} \colon
\begin{cases}
\, \mathbb{R}_{> 0} \times (0, \pi) \times [0, 2 \pi) \rightarrow \mathbb{R}_{> 0} \times (- 1, 1) \times [0, 2 \pi)  \\[0.25cm]
\hspace{2.16cm} (r, \vartheta, \phi) \mapsto (x_1, x_2, x_3) 
\end{cases} 
\end{equation*}
with
\begin{equation*}
x_1 = \frac{r^3}{3} \, , \quad x_2 = - \cos{(\vartheta)} \, , \quad \textnormal{and} \quad x_3 = \phi \, .
\end{equation*}
Expressed in these coordinates, the metric ansatz in Equation (\ref{MetricAnsatz}) becomes
\begin{equation*} 
\boldsymbol{g} = F(r) \, \textnormal{d}t \otimes \textnormal{d}t - \frac{1}{r^2} \biggl(\frac{G(r)}{r^2} + H(r)\biggr) \textnormal{d}x_1 \otimes \textnormal{d}x_1 - G(r) \, r^2 \biggl[\frac{\textnormal{d}x_2 \otimes \textnormal{d}x_2}{1 - x_2^2} + \bigl(1 - x_2^2\bigr) \, \textnormal{d}x_3 \otimes \textnormal{d}x_3\biggr] \, ,
\end{equation*}
where the radial distance in unimodular spherical coordinates is $r = (3 x_1)^{1/3}$. Using the abbreviations $f_1 = r^{- 2} \, [G/r^2 + H]$, $f_2 = f_3 = G r^2$, and $f_4 = F$,\footnote{To meet the assumptions Schwarzschild imposed on the metric $\boldsymbol{g}$, these functions have to fulfill the following set of conditions:
\begin{itemize}
\item[1.] In the limit $x_1 \rightarrow \infty$, $f_1 = r^{- 4} = (3 x_1)^{- 4/3}$, $f_2 = f_3 = r^2 = (3 x_1)^{2/3}$, and $f_4 = 1$.
\item[2.] The determinantial equation $f_1 f_2 f_3 f_4 = 1$ has to be satisfied.
\item[3.] The metric formed by the tuple $(f_i)_{i \in \{1, 2, 3, 4\}}$ is an exact solution to the Einstein field equations (\ref{UnimodularEFE}).
\item[4.] All $f_i$, $i \in \{1, 2, 3, 4\}$, have to be continuous except at $x_1 = 0$.
\end{itemize} \label{fCond}} 
and inserting this ansatz into the Einstein field equations (\ref{UnimodularEFE}), Schwarzschild finds the solutions
\begin{equation} \label{MetricalCoeffs} 
f_1 = \frac{(3 x_1 + \alpha^3)^{- 1}}{\bigl[(3 x_1 + \alpha^3)^{1/3} - \alpha\bigr]} \, , \quad f_2 = f_3 = (3 x_1 + \alpha^3)^{2/3} \, , \quad \textnormal{and} \quad f_4 = 1 - \frac{\alpha}{(3 x_1 + \alpha^3)^{1/3}} \, ,
\end{equation}
in which $\alpha \in \mathbb{R}_{> 0}$ is a constant of integration. Finally, to further simplify the metric representation corresponding to these solutions, he employs the transformation from the unimodular spherical polar coordinates to what are today called Schwarzschild coordinates
\begin{equation*} 
\mathfrak{T}^{(3)} \colon
\begin{cases}
\, \mathbb{R}_{> 0} \times (- 1, 1) \times [0, 2 \pi) \rightarrow (\alpha, \infty) \times (0, \pi) \times [0, 2 \pi)  \\[0.25cm]
\hspace{1.89cm} (x_1, x_2, x_3) \mapsto (R, \vartheta, \phi) 
\end{cases} 
\end{equation*}
with
\begin{equation*} 
R = (3 x_1 + \alpha^3)^{1/3} \, , \quad \vartheta = \pi - \arccos{(x_2)} \, , \quad \textnormal{and} \quad \phi = x_3 \, .
\end{equation*}
This yields the familiar expression for the metric accounting for the gravitational field of a massive, spherically symmetric point particle, 
\begin{equation} \label{SchwarzschildRep} 
\boldsymbol{g} = \biggl(1 - \frac{\alpha}{R}\biggr) \textnormal{d}t \otimes \textnormal{d}t - \biggl(1 - \frac{\alpha}{R}\biggr)^{- 1} \textnormal{d}R \otimes \textnormal{d}R - R^2 \boldsymbol{g}_{S^2} \, ,
\end{equation}
where $\boldsymbol{g}_{S^2} = \textnormal{d}\vartheta \otimes \textnormal{d}\vartheta + \sin^2{(\vartheta)} \, \textnormal{d}\phi \otimes \textnormal{d}\phi$ is the metric on the unit $2$-sphere.\footnote{According to Birkhoff's theorem (see, e.g., \cite[Appendix B]{HawkingEllis}), this so-called exterior Schwarzschild solution of the vacuum Einstein field equations is unique on the specified coordinate range. More precisely, the theorem states that any $C^2$ vacuum solution of the Einstein field equations that is spherically symmetric in an open set $U$ is locally isometric to the (maximally analytically extended) Schwarzschild solution in $U$. Hence, the exterior Schwarzschild solution may also be used to describe \textit{dynamical} scenarios as long as spherical symmetry is conserved and the dynamical process under consideration does not give rise to violations of the validity of the vacuum Einstein field equations, that is, changes of the volume of the source have to be radially symmetrically decreasing in nature and not expand the source into the original vacuum region.}
\begin{Remark}
In \cite{Schwarzschild1}, Schwarzschild merely states that the constant of integration $\alpha$ depends on the magnitude of the mass of the gravitational-field-generating point particle. It was only later explicitly identified with the constant $2 M$ (given in natural units), where $M \in \mathbb{R}_{> 0}$ is the common Newtonian mass of a massive point particle.\footnote{The parameter $M$ can furthermore be shown to coincide with the ADM mass of the point particle (for details on the ADM mass see, e.g., \cite[Chapter 11.2]{Wald}). Besides, extending the mass to negative values $M \in \mathbb{R}_{< 0}$, the exterior Schwarzschild solution contains a naked singularity.} For this purpose, one usually takes the weak-field limit, in which $\alpha \ll R$,\footnote{Here, the weak-field limit applies for either movements at large distances from the gravitating source or for extremely light sources.} of the geodesic equation of motion, thus making it concordant with Newton's law of motion for a test particle moving in the Newtonian gravitational field generated by a massive point particle. This comparison determines the weak-field limit $g_{t t} \simeq 1 - 2 M/R$ of the temporal component of the metric (\ref{SchwarzschildRep}), and therefore the constant of integration to be $\alpha = 2M$. In the limit $M \rightarrow 0$, one expectedly obtains the Minkowski metric expressed in spherical polar coordinates.
\end{Remark}
\begin{Remark}
For the given choice of time coordinate $t$, the metric representation (\ref{SchwarzschildRep}) is invariant under both time translations $t \mapsto t' = t + \textnormal{const.}$ and time reversals $t \mapsto t' = - t$. It is further invariant under rotational and reflective transformations of the angular coordinates $(\vartheta, \phi)$ about the origin. Moreover, Schwarzschild's radial coordinate $R$, which is a generalized spherical polar-type coordinate that behaves like the usual spherical polar radial coordinate $r$ only in the asymptotic end (i.e., for $R \rightarrow \infty$), has the property that each sphere $\bigl\{(t, R, \vartheta, \phi) \, \big| \, t, R = \textnormal{const.}, \vartheta \in (0, \pi), \,\, \textnormal{and} \,\, \phi \in [0, 2 \pi)\bigr\}$ has an intrinsic surface area of $4 \pi R^2$. 
\end{Remark}
\begin{Remark} \label{RemLocPMS}
By nonregularly extending the Schwarzschild representation (\ref{SchwarzschildRep}) of the exterior Schwarzschild solution across the hypersurface at $R = \alpha$ into the (nonstatic) interior region with $\textnormal{Ran}(R) = (0, \alpha)$, where it still satisfies the vacuum Einstein field equations (\ref{UnimodularEFE}) and Conditions ($\mathcal{C}1$)--($\mathcal{C}3$), one finds, in addition to the coordinate singularity that already exists in a limiting sense at $R = \alpha$, a curvature singularity at the location $R = 0$, which, however, corresponds to $r = - \alpha$, not coinciding with the by Schwarzschild originally assumed massive point particle location at $r = 0$. Accordingly, following the usual placeholder interpretation that identifies the gravitational-field-generating source with the curvature singularity at $R = 0$, implying the complete arbitrariness of coordinates, Schwarzschild assigning the location of the point particle to $r = 0$ in the first place was unfounded. Nevertheless, at that time, it was not yet known that the radial location at $r = 0$ does not correspond to the true origin of the Schwarzschild solution, i.e., the curvature singularity, but rather to the location of its event horizon.\footnote{The (future component of the) event horizon is a trapping null hypersurface for which observers stationed at rest at infinity find that there exists no infalling (physical) matter--energy distribution in the Universe that can ever, as a direct consequence of gravitational time dilation, traverse it. While closing in on the event horizon, every matter--energy distribution seen by those observers also undergoes a gravitational redshift that eventually leads to a full dimming at the horizon. For comoving observers, on the other hand, infalling matter--energy distributions may cross the event horizon unidirectionally in finite eigentime into a region of spacetime in which an inexorable dragging towards the curvature singularity at $R = 0$ occurs. Framed mathematically more rigorously, the future component of the event horizon is the future boundary of the causal past of future null infinity, $\partial J^-(\mathscr{I}^+) \cap \mathfrak{M}$.}$^{,}$\footnote{The concurrence of the spherical polar coordinate singularity at $r = 0$ and the location of the event horizon is a direct consequence of Schwarzschild relating the former, and thus the supposed location of the source, to the location of the pole of the metrical coefficient $f_1$ in Equation (\ref{MetricalCoeffs}) (which he refers to as a ``point of discontinuity'') by suitably arranging the values of the two constants of integration in his solution to the vacuum Einstein field equations, for he requires his solution to satisfy the continuity condition 4.\ specified in Footnote \ref{fCond}.} 

The curvature singularity being at $R = 0$ also shows that the nonregularly extended Schwarzschild coordinate representation is distinguished in so far as it allows the gravitational-field-generating source to be considered a massive point particle located at the origin. In contrast, for coordinate systems with radial coordinates of the form, say, $R' = R + \beta$, where $\beta \in \mathbb{R}_{> 0}$, the origin of the Schwarzschild radial coordinate $R$ is ``inflated'' into a sphere of finite, nonzero radius $R' = \beta$, rendering the shape of the source a $2$-sphere instead of a point. However, both of these source representations are equivalent in that they have the same center of mass, and therefore yield the same gravitational effects. In other words, the gravitational effects of a point source at $R = 0$ and a spherical source with radius $R' = \beta$ are indistinguishable.
\end{Remark}

\subsection{Modern Account of the Exterior Schwarzschild Solution and Kruskal's Maximal Analytic Extension} \label{SubsectionIIB}

\noindent From a more modern point of view, the exterior Schwarzschild solution $\boldsymbol{g}$ is extended to the pair $(\mathfrak{M}, \boldsymbol{g})$, called the exterior Schwarzschild spacetime, which is a smooth, connected, globally hyperbolic and asymptotically flat Lorentzian $4$-manifold homeomorphic to $\mathbb{R}^2 \times S^2$, where the metric $\boldsymbol{g}$ defined in Equation (\ref{SchwarzschildRep}) in Schwarzschild coordinates $(t, r, \vartheta, \phi) \in \mathbb{R} \times (2 M, \infty) \times (0, \pi) \times [0, 2 \pi)$ is a $1$-parameter family of exact, spherically symmetric and static Petrov type D solutions of the vacuum Einstein field equations $\textnormal{Ric}(\boldsymbol{g}) = \boldsymbol{0}$.\footnote{Here and in the following, we use the more common symbol $r$ for the Schwarzschild radial coordinate $R$.}$^{,}$\footnote{For details on the causal structure of spacetimes and the Petrov classification of the algebraic symmetries of the Weyl tensor see, e.g., \cite{Minguzzi, MinguzziSanchez} and \cite[Chapter 4]{SKMHH}, respectively.} Extending this spacetime across the (null) hypersurface at $r = 2 M$ up to the origin at $r = 0$, which in Schwarzschild coordinates is a nonregular extension with a coordinate singularity at $r = 2 M$, the spacetime now consists of two connected components, namely the exterior component $\textnormal{B}_{\textnormal{I}} = \bigl\{(t, r, \vartheta, \phi) \, \big| \, t \in \mathbb{R}, \, r \in (2 M, \infty), \, \vartheta \in (0, \pi), \, \phi \in [0, 2 \pi)\bigr\}$, which is the domain of outer communication, and the interior component $\textnormal{B}_{\textnormal{II}} = \bigl\{(t, r, \vartheta, \phi) \, \big| \, t \in \mathbb{R}, \, r \in (0, 2 M), \, \vartheta \in (0, \pi), \, \phi \in [0, 2 \pi)\bigr\}$, which is the (future) trapped region.\footnote{In other words, $\textnormal{B}_{\textnormal{I}}$ and $\textnormal{B}_{\textnormal{II}}$ are two disjoint regions separated by a singular hypercylinder at $r = 2 M$.} Then, the Schwarzschild metric expressed in Schwarzschild coordinates is defined for all $r \in \mathbb{R}_{> 0} \backslash \{2 M\}$, nonstatic in the interior region $\textnormal{B}_{\textnormal{II}}$, and features two types of singularities, viz., a curvature singularity at $r = 0$,\footnote{The well-known fact of the inevitability of an infalling observer impinging onto this particular curvature singularity after having crossed the event horizon can be directly traced back to the singularity's spacelike nature.} and the above-mentioned coordinate singularity at $r = 2 M$.
\begin{Remark} \label{SchTime}
The Schwarzschild time coordinate $t$ is a Cauchy temporal function in the exterior region $\textnormal{B}_{\textnormal{I}}$, i.e., $t \colon \textnormal{B}_{\textnormal{I}} \rightarrow \mathbb{R}$ is a smooth function with future-directed, timelike gradient $\boldsymbol{\nabla} t = g^{\mu \nu} (\partial_{\mu} t) \partial_{\nu} = (1 - 2 M/r)^{- 1} \, \partial_t$ for which the level sets $t^{- 1}(\, . \,)$ that foliate this region are smooth, spacelike Cauchy hypersurfaces (see, e.g., \cite{BernalSanchez}). Since $t$ is not timelike in the interior region $\textnormal{B}_{\textnormal{II}}$ and behaves irregularly at $r = 2 M$, its level sets do not, however, form a regular foliation of the extension $\textnormal{B}_{\textnormal{I}} \cup \textnormal{B}_{\textnormal{II}}$.
\end{Remark}

\noindent As our later focus with regard to the construction procedure of Penrose diagrams is on the maximal analytic extension of the exterior Schwarzschild spacetime, the so-called Kruskal extension,\footnote{The Kruskal extension is also the maximal Cauchy development of the exterior Schwarzschild spacetime. It has, except for the metric not being globally static anymore, the same geometric and topological characteristics as the exterior Schwarzschild spacetime itself.} we now summarize the main result of Kruskal's seminal 1960 paper \cite{Kruskal} on the topic. In this paper, Kruskal demonstrates that the then so-called Schwarzschild singularity of the exterior Schwarzschild solution occurring at $r = 2 M$ is merely apparent, and therefore not an actual singularity that has a physical effect. To this end, he employs a ``seemingly simpler and more explicit'' choice of coordinates (as compared to the coordinates used in the works of Kasner \cite{Kasner}, Lema\^{i}tre \cite{Lemaitre}, Einstein and Rosen \cite{EinsteinRosen}, Robertson \cite{Robertson}, Synge \cite{Synge}, Finkelstein \cite{Finkelstein}, and Fronsdal \cite{Fronsdal}) in which the Schwarzschild singularity is removed, and which gives rise to the maximal analytic extension of Schwarzschild's original solution.\footnote{Other very influential coordinate systems for the exterior Schwarzschild solution at that time, which are not mentioned by Kruskal, are, e.g., the ones by Flamm \cite{Flamm}, Gullstrand \cite{Gullstrand}, Painlev\'e \cite{Painleve1, Painleve2}, and Weyl \cite{Weyl}.}$^{,}$\footnote{Although Lema\^{i}tre, already in the mid 1930s, had explicitly demonstrated in \cite{Lemaitre} that the Schwarzschild singularity is only a fictitious coordinate singularity, most researchers were convinced that it is a physical singularity until the end of the 1950s (for a possible reason see Remark \ref{RemLocPMS}).}$^{,}$\footnote{Around the same time, Szekeres published the article \cite{Szekeres} in which he presents essentially the same coordinate system as Kruskal (except for a different scaling) to also show that the Schwarzschild singularity at $r = 2 M$, which he calls a ``Schwarzschild hypercylinder,'' is merely an apparent singularity removable by a suitable change of coordinates. In the same article, Szekeres furthermore provided one of the first systematic, coordinate-independent definitions of singularities of pseudo-Riemannian manifolds.}

More precisely, Kruskal starts with the exterior Schwarzschild solution (\ref{SchwarzschildRep}) expressed in the usual Schwarzschild coordinates $(t, r, \vartheta, \phi) \in \mathbb{R} \times (2 M, \infty) \times (0, \pi) \times [0, 2 \pi)$, albeit with signature $(3, 1, 0)$, and seeks coordinates $(v, u, \vartheta', \phi') \in \mathbb{R} \times \mathbb{R}_{> 0} \times (0, \pi) \times [0, 2 \pi)$ in which the metric is still spherically symmetric but radial light rays have slopes $\textnormal{d}u/\textnormal{d}v = \pm 1$ \textit{everywhere}, so that the light cone structure at each event agrees with that of Minkowski spacetime.\footnote{This property avoids the pathologies of the local light cone structure at the event horizon occurring in the Schwarzschild coordinate representation (see Section \ref{SubsectionIVA} for more details) and renders the global causal structure most transparent.}$^{,}$\footnote{For historical accuracy, we here employ Kruskal's original coordinate terminology, that is, instead of the well-established present-day designators $T$ and $X$ for the Kruskal time and radial coordinates, respectively, we adopt the notation $v$ and $u$. It is, however, paramount that Kruskal's notation is not to be confused with the standard notation for the various Eddington--Finkelstein coordinates below.} Since these two conditions immediately lead to the metric ansatz
\begin{equation*}
\boldsymbol{g} = f(r)^2 \, [- \textnormal{d}v \otimes \textnormal{d}v + \textnormal{d}u \otimes \textnormal{d}u] + r^2 \boldsymbol{g}_{S^2} \, ,
\end{equation*}
a direct comparison with Schwarzschild's metric representation, requiring that the function $f$ is finite and nonzero for $v = u = 0$, results in the coordinate transformation
\begin{equation*} 
\mathfrak{T}^{(\textnormal{K})} \colon
\begin{cases}
\, \mathbb{R} \times (2 M, \infty) \times (0, \pi) \times [0, 2 \pi) \rightarrow \mathbb{R} \times \mathbb{R}_{> 0} \times (0, \pi) \times [0, 2 \pi)  \\[0.25cm]
\hspace{3.27cm} (t, r, \vartheta, \phi) \mapsto (v, u, \vartheta', \phi') 
\end{cases} 
\end{equation*}
with
\begin{equation*}
v = \sqrt{\frac{r}{2 M} - 1} \,\, e^{r/(4 M)} \sinh{\biggl(\frac{t}{4 M}\biggr)} \, , \quad u = \sqrt{\frac{r}{2 M} - 1} \,\, e^{r/(4 M)} \cosh{\biggl(\frac{t}{4 M}\biggr)} \, , \quad \vartheta' = \vartheta \, , \quad \textnormal{and} \quad \phi' = \phi
\end{equation*}
for $|v| < u$, on the one hand,\footnote{The particular restriction on the ranges of the Kruskal coordinates $v$ and $u$ in this region originates directly from the respective range of the radial coordinate $r$ according to Equation (\ref{ruv-eq}) below.} and the so-called Kruskal representation of the exterior Schwarzschild solution
\begin{equation} \label{KRSM} 
\boldsymbol{g} = \frac{32 M^3 \, e^{- r(v, u)/(2 M)}}{r(v, u)} \, [- \textnormal{d}v \otimes \textnormal{d}v + \textnormal{d}u \otimes \textnormal{d}u] + r^2 \boldsymbol{g}_{S^2} \, , 
\end{equation}
on the other hand. Here, the function $r(v, u)$ is a solution of the transcendental algebraic equation 
\begin{equation} \label{ruv-eq}
\biggl(\frac{r}{2 M} - 1\biggr) e^{r/(2 M)} = u^2 - v^2	\, , 
\end{equation}
namely $r(v, u) = 2 M \bigl[1 + W_0\bigl(e^{- 1} [u^2 - v^2]\bigr)\bigr]$, with $W_0$ being the principal branch of the Lambert $W$ function. The Kruskal representation (\ref{KRSM}) of the Schwarzschild solution, which is, strictly speaking, only valid for $r > 2 M$, may, however, be analytically extended across the hypersurface at $r = 2 M$ without the Schwarzschild singularity ever arising. This makes it regular for all $r \in \mathbb{R}_{> 0}$.
\begin{Remark} \label{SchwInt}
Even though the metric is of the same form when $r < 2 M$, the transformation from Schwarzschild coordinates to Kruskal coordinates in the extending region reads differently, viz.,
\begin{equation*} 
\mathfrak{T}^{(\textnormal{K})} \colon
\begin{cases}
\, \mathbb{R} \times (0, 2 M) \times (0, \pi) \times [0, 2 \pi) \rightarrow \mathbb{R}_{> 0} \times \mathbb{R} \times (0, \pi) \times [0, 2 \pi) \\[0.25cm]
\hspace{3.10cm} (t, r, \vartheta, \phi) \mapsto (v, u, \vartheta', \phi') 
\end{cases} 
\end{equation*}
with
\begin{equation*}
v = \sqrt{1 - \frac{r}{2 M}} \,\, e^{r/(4 M)} \cosh{\biggl(\frac{t}{4 M}\biggr)} \, , \quad u = \sqrt{1 - \frac{r}{2 M}} \,\, e^{r/(4 M)} \sinh{\biggl(\frac{t}{4 M}\biggr)} \, , \quad \vartheta' = \vartheta \, , \quad \textnormal{and} \quad \phi' = \phi 
\end{equation*}
for $|u| < v < \sqrt{1 + u^2}$. Nonetheless, the function $r(v, u)$ is still determined by Equation (\ref{ruv-eq}), also yielding $r(v, u) = 2 M \bigl[1 + W_0\bigl(e^{- 1} [u^2 - v^2]\bigr)\bigr]$.
\end{Remark}

\noindent We point out that the \textit{maximal} analytic extension of the Kruskal representation of the exterior Schwarzschild solution comprises \textit{two} images of the region $\bigl\{(t, r, \vartheta, \phi) \, \big| \, t \in \mathbb{R}, \, r \in (2 M, \infty), \, \vartheta \in (0, \pi), \, \phi \in [0, 2 \pi)\bigr\}$ exterior to the Schwarzschild singularity,\footnote{Hence, the Kruskal extension possesses two asymptotic ends.} that is,
\begin{equation*}
\begin{split}
\textnormal{B}_{\textnormal{I}} & = \bigl\{(v, u, \vartheta', \phi') \, \big| \, v \in \mathbb{R}, \, u \in \mathbb{R}_{> 0}, \, \vartheta' \in (0, \pi), \, \phi' \in [0, 2 \pi) \,\,\, \textnormal{and} \,\,\, |v| < u\bigr\} \\[0.2cm]
\textnormal{B}_{\textnormal{III}} & = \bigl\{(v, u, \vartheta', \phi') \, \big| \, v \in \mathbb{R}, \, u \in \mathbb{R}_{< 0}, \, \vartheta' \in (0, \pi), \, \phi' \in [0, 2 \pi) \,\,\, \textnormal{and} \,\,\, |v| < - u\bigr\} \, ,
\end{split}
\end{equation*}
and \textit{two} images of the interior region $\bigl\{(t, r, \vartheta, \phi) \, \big| \, t \in \mathbb{R}, \, r \in (0, 2 M), \, \vartheta \in (0, \pi), \, \phi \in [0, 2 \pi)\bigr\}$, namely 
\begin{equation*}
\begin{split}
\textnormal{B}_{\textnormal{II}} & = \bigl\{(v, u, \vartheta', \phi') \, \big| \, v \in \mathbb{R}_{> 0}, \, u \in \mathbb{R}, \, \vartheta' \in (0, \pi), \, \phi' \in [0, 2 \pi) \,\,\, \textnormal{and} \,\,\, |u| < v < \sqrt{1 + u^2}\bigr\} \\[0.2cm]
\textnormal{B}_{\textnormal{IV}} & = \bigl\{(v, u, \vartheta', \phi') \, \big| \, v \in \mathbb{R}_{< 0}, \, u \in \mathbb{R}, \, \vartheta' \in (0, \pi), \, \phi' \in [0, 2 \pi) \,\,\, \textnormal{and} \,\, - \sqrt{1 + u^2} < v < - |u|\bigr\} \, ,
\end{split}
\end{equation*}
where the regions $\textnormal{B}_{\textnormal{III}}$ and $\textnormal{B}_{\textnormal{IV}}$ are obtained from the regions $\textnormal{B}_{\textnormal{I}}$ and $\textnormal{B}_{\textnormal{II}}$, respectively, by simultaneously applying the discrete reversal isometries $v \mapsto v' = - v$ and $u \mapsto u' = - u$, extending the range of the Kruskal radial coordinate $u$ in region $\textnormal{B}_{\textnormal{I}}$ from $\mathbb{R}_{> 0}$ to $\mathbb{R}_{< 0}$, and similarly for the Kruskal time coordinate $v$ in region $\textnormal{B}_{\textnormal{II}}$. Kruskal proves argumentatively that this extension is indeed maximal by analyzing the global behavior of geodesics.\footnote{More precisely, he remarks that in this analytic extension, geodesics either run into the ``barrier'' of intrinsic singularities at $r = 0$ or can be continued indefinitely with respect to their ``natural length,'' precluding the existence of an even larger extension.}

\section{Central Idea Behind Penrose Diagrams} \label{SectionIII}

\noindent Before we explicitly construct and then visualize the Penrose diagram of the maximally analytically extended Schwarzschild spacetime, we present the central idea underlying---and the general framework combined with a precise mathematical definition for---Penrose diagrams.\footnote{Details on the background and the construction of Penrose diagrams can also be found in, e.g., \cite{SchindlerAguirre, ChruscielÖlzSzybka, DafermosRodnianski1, DafermosRodnianski2}.} To this end, we let $(\mathfrak{M}, \boldsymbol{g})$ be a smooth, connected, and time-oriented Lorentzian $4$-manifold. In order to obtain a faithful $2$-dimensional conformally Minkowskian representation of this manifold that comprises essential information on the global $4$-dimensional causal structure and on possible infinities, so that specific contextual causal relations between points in the Lorentzian $4$-manifold can be accurately and most transparently depicted on a finite-sized $2$-dimensional diagram (where these relations always correspond to particular relations between points in the diagram), one has to, in a first step, either suitably project out two of the four dimensions or just consider a $2$-dimensional restriction of the Lorentzian $4$-manifold. With the exception of the distinguished restrictions to totally geodesic submanifolds, where every geodesic is also a geodesic when regarded in the full Lorentzian $4$-manifold (cf.\ Remark \ref{Kerr} for the example of the nonextreme Kerr spacetime), different restrictions of one and the same spacetime can lead to possibly different Penrose diagrams each containing information intrinsic only to a particular slice. Therefore, it may in general seem favorable to work with projections (where every point is identified with the entire ``extracted'' $2$-dimensional submanifold at that point), for then, although different projections may also result in different Penrose diagrams depicting different aspects of the same global $4$-dimensional causal structure, they always contain information characteristic of the \textit{full} Lorentzian $4$-manifold. However, the nature of these projections must be such that they always \textit{conformally} map onto the representative $2$-dimensional submanifolds, while at the same time they project out \textit{preferably compact} $2$-dimensional submanifolds. The underlying reason for why conformal projections are needed is that, even though they are in general not isometric so that scales may overall change, they are locally angle- and orientation-preserving. This brings about the necessary property that under their action the local geometry, and thus the local causal structure, remains invariant, i.e., null geodesics in the $2$-dimensional representative submanifold correspond to a certain pivotal subset of null geodesics in the original Lorentzian $4$-manifold. But since this type of projection does not exist for Lorentzian $4$-manifolds of all geometric and topological shapes, one needs to first identify families of Lorentzian $4$-manifolds that indeed admit projections preserving the $4$-dimensional conformal structure in the above sense.\footnote{Although spacetime symmetries are particularly helpful in finding such projections, they are, similar to the conditions of asymptotic and conformal flatness, not necessary for the construction of Penrose diagrams.} The simplest example of one such family is given by the so-called warped product spacetimes that consist of a product space $\mathfrak{M} = \mathfrak{M}^{(2)}_{\textnormal{L}} \times \mathfrak{M}^{(2)}_{\textnormal{R}}$ endowed with a metric of the form
\begin{equation} \label{prodstr}
\boldsymbol{g} = \boldsymbol{g}^{(2)}_{\textnormal{L}} \oplus \bigl(f \hspace{0.02cm} \boldsymbol{g}^{(2)}_{\textnormal{R}}\bigr) \quad \textnormal{for} \quad f \colon \mathfrak{M}^{(2)}_{\textnormal{L}} \rightarrow \mathbb{R}_{> 0} \, ,
\end{equation}
where $(\mathfrak{M}^{(2)}_{\textnormal{L}}, \boldsymbol{g}^{(2)}_{\textnormal{L}})$ and $(\mathfrak{M}^{(2)}_{\textnormal{R}}, \boldsymbol{g}^{(2)}_{\textnormal{R}})$ are $2$-dimensional Lorentzian and Riemannian manifolds, respectively \cite[Chapter 7]{ONeill2}. As this product structure allows for the natural identification 
\begin{equation*}
T\bigl(\mathfrak{M}^{(2)}_{\textnormal{L}} \times \mathfrak{M}^{(2)}_{\textnormal{R}}\bigr) = T\mathfrak{M}^{(2)}_{\textnormal{L}} \oplus T\mathfrak{M}^{(2)}_{\textnormal{R}} \, ,
\end{equation*}
one obtains the splitting
\begin{equation*}
\boldsymbol{g}(\boldsymbol{Y}_{\textnormal{L}} + \boldsymbol{Y}_{\textnormal{R}}, \boldsymbol{Z}_{\textnormal{L}} + \boldsymbol{Z}_{\textnormal{R}}) = \boldsymbol{g}^{(2)}_{\textnormal{L}}(\boldsymbol{Y}_{\textnormal{L}}, \boldsymbol{Z}_{\textnormal{L}}) + f \hspace{0.02cm} \boldsymbol{g}^{(2)}_{\textnormal{R}}(\boldsymbol{Y}_{\textnormal{R}}, \boldsymbol{Z}_{\textnormal{R}}) 
\end{equation*}
for $\boldsymbol{Y}_k, \boldsymbol{Z}_k \in \Gamma\bigl(T\mathfrak{M}^{(2)}_k\bigr)$, $k \in \{\textnormal{L}, \textnormal{R}\}$, so any null geodesic in the $2$-dimensional Lorentzian manifold $(\mathfrak{M}^{(2)}_{\textnormal{L}}, \boldsymbol{g}^{(2)}_{\textnormal{L}})$ is also a null geodesic in the full Lorentzian $4$-manifold $(\mathfrak{M}, \boldsymbol{g})$ for a fixed Riemannian component $(\mathfrak{M}^{(2)}_{\textnormal{R}}, \boldsymbol{g}^{(2)}_{\textnormal{R}})$.\footnote{For the special case of spherically symmetric spacetimes, one finds that $\mathfrak{M}^{(2)}_{\textnormal{L}} \cong \mathfrak{M} \slash \textnormal{SO}(3)$ and $\mathfrak{M}^{(2)}_{\textnormal{R}} \cong S^2$, that is, every point of the associated Penrose diagram corresponds to a $2$-sphere. Accordingly, in the diagram, the global $4$-dimensional radial light cone structure is fully preserved. Moreover, there is no practical difference between radial projections and radial restrictions for these spacetimes.} Certain causal relations between different points in $(\mathfrak{M}, \boldsymbol{g})$ can thus be simply analyzed by means of the causal relations between points in $(\mathfrak{M}^{(2)}_{\textnormal{L}}, \boldsymbol{g}^{(2)}_{\textnormal{L}})$. (For a precise mathematical definition of the procedure of reducing the original Lorentzian $4$-manifold to a representative $2$-dimensional Lorentzian submanifold by projecting, where projecting is understood as every smooth timelike curve in the submanifold being a projection of a smooth timelike curve in the full manifold, which, under certain causality conditions on the spacetimes, ensures the invariance of causal relations and ultimately leads to a new type of $2$-dimensional diagram coined \textit{projection diagram}, see \cite{ChruscielÖlzSzybka}.)

\begin{Remark}
Since the Lorentzian manifold $(\mathfrak{M}^{(2)}_{\textnormal{L}}, \boldsymbol{g}^{(2)}_{\textnormal{L}})$ is of dimension two, it is inherently conformally flat, i.e., it is conformally equivalent\footnote{Two metrics $\boldsymbol{g}$ and $\boldsymbol{g}'$ are conformally equivalent if and only if $\boldsymbol{g}' = \lambda \boldsymbol{g}$ for $\lambda \in C^{\infty}(\mathfrak{M}, \mathbb{R}_{> 0})$. \label{ConfEquiv}} to a subset of Minkowski spacetime $\mathbb{R}^{1, 1}$ with its particular causal structure.\footnote{Throughout the paper, we denote the Minkowski spacetime $(\mathbb{R}^n, \boldsymbol{\eta}^{(n)})$ of total dimension $n$ by $\mathbb{R}^{1, n - 1}$.} 
\end{Remark}
\begin{Remark} \label{Kerr}
The maximally analytically extended Schwarzschild spacetime discussed in Section \ref{SubsectionIIB} is a simple example of a warped product spacetime for which the function $f$ in Equation (\ref{prodstr}) is given by the square of the radial distance $r$ [cf.\ Equation (\ref{KRSM})]. But already its canonical rotating generalization, the nonextreme Kerr spacetime (for details see, e.g., \cite[Chapter 6]{ChandraBook} and \cite{ONeill1}), does not possess the particular product structure of warped product spacetimes anymore, which is due to nonremovable cross terms in the metric that result from the presence of nonzero angular momentum generating rigid rotations around the symmetry axis. This aspect renders the realization of the above type of conformal projection impossible for this spacetime. Nonetheless, it is still useful to draw Penrose diagrams for $2$-dimensional submanifolds that are, as initially stated, totally geodesic restrictions, such as the axis of symmetry with a polar angular coordinate $\vartheta \in \{0, \pi\}$ \cite{Carter} or the equatorial plane at $\vartheta = \pi/2$ at an arbitrary but fixed value of the azimuthal angular coordinate $\phi \in [0, 2 \pi)$ \cite{ChruscielÖlzSzybka}, for then every geodesic in the $2$-dimensional representative submanifolds is also a geodesic in the full nonextreme Kerr spacetime. 
\end{Remark}

After having singled out a suitable $2$-dimensional submanifold representative of the Lorentzian $4$-manifold, one performs, in a second step, a globally regular compactification that conformally maps the entire (possibly infinite) submanifold into a region of finite size with infinity on the boundary,\footnote{This particular type of compactification is, similar to the conformal projections, locally angle- and orientation-preserving, but does not necessarily preserve distances.}$^{,}$\footnote{The process of regular, conformal compactification may involve gluings of in a certain sense maximal, disjoint sets that ultimately make up the spacetime regions to be visualized. For more details see Remark \ref{GCP}.} bringing about the advantage that one can encode and visualize essential aspects of the global $4$-dimensional geometry and causal structure, particularly the light cone structure and infinity, in a finite-sized $2$-dimensional conformally Minkowskian diagrammatic representation, the so-called Penrose diagrams. We can define both the regular, conformal compactification and Penrose diagrams mathematically more rigorously as follows.
\begin{Def} \label{DefPD}
Let $(\mathfrak{M}^{(2)}, \boldsymbol{g}^{(2)})$ be a smooth, connected, and time-oriented Lorentzian $2$-manifold, possibly with a boundary, and $\mathcal{U} \subset \mathfrak{M}^{(2)}$ an open subset. Furthermore, let $\Phi \colon \mathcal{U} \rightarrow \mathbb{R}^2$ be a diffeomorphism for which
\begin{itemize}
\item[($\mathcal{C}1$)] the closure $\overline{\mathcal{U}} = \mathfrak{M}^{(2)}$.
\item[($\mathcal{C}2$)] there exists a compact subset $\mathcal{V} \subset \mathbb{R}^2$ such that the image $\Phi(\mathcal{U}) \subset \mathcal{V}$.
\item[($\mathcal{C}3$)] the metric takes the form $\boldsymbol{g}^{(2)} = \lambda \boldsymbol{\eta}^{(2)}$, where $\lambda \in C^{\infty}(\mathcal{U}, \mathbb{R}_{> 0})$ and $\boldsymbol{\eta}^{(2)}$ is the $2$-dimensional Minkowski metric.
\end{itemize}
A Penrose diagram is the pair $(\mathcal{U}, \Phi)$, with the image of $\Phi(\mathcal{U})$ being called the domain of the Penrose diagram and the boundary of the closure of the image of $\Phi(\mathcal{U})$ referred to as conformal infinity of $\mathfrak{M}^{(2)}$.
\end{Def}
\noindent Here, Conditions ($\mathcal{C}1$) and ($\mathcal{C}2$) ensure that the Penrose diagram comprises the entirety of $\mathfrak{M}^{(2)}$ in a finite-sized representation, which, in particular, allows for a proper study of conformal infinity. Condition ($\mathcal{C}3$) prompts $\Phi(\mathcal{U})$ to be a conformal isometry from $(\mathcal{U}, \boldsymbol{g}^{(2)})$ into a bounded subset $(\mathcal{V}, \boldsymbol{\eta}^{(2)}) \subset \mathbb{R}^{1, 1}$ of $2$-dimensional Minkowski spacetime, so that the causal structure of $(\mathcal{U}, \boldsymbol{g}^{(2)})$ is preserved everywhere in $(\mathcal{V}, \boldsymbol{\eta}^{(2)})$.

\section{Construction Procedure of the Penrose Diagram of the Maximally Analytically Extended Schwarzschild Spacetime} \label{SectionIV}

\subsection{Coordinate Representation for the Penrose Diagram} \label{SubsectionIVA}

\noindent Having outlined the central idea and the general framework, we now come to the construction procedure of Penrose diagrams demonstrated through the example of the maximally analytically extended Schwarzschild spacetime. For this purpose, we first recall the derivation of the canonical variant of compactified Kruskal--Szekeres spacetime coordinates, which not just cover the entire maximal analytic extension in a regular manner, but also constitute a suitable basis for the corresponding Penrose diagram as they give rise to a metric representation that is of the particular warped product structure specified in Equation (\ref{prodstr}) and satisfy Conditions ($\mathcal{C}1$)--($\mathcal{C}3$) in Definition \ref{DefPD}. To explain in detail the reasoning behind \textit{all} the steps involved in this derivation,\footnote{The steps involved are, in essence, a series of particular coordinate transformations.} our starting point is the exterior Schwarzschild metric (\ref{SchwarzschildRep}) expressed in Schwarzschild coordinates $(t, r, \vartheta, \phi) \in \mathbb{R} \times (2 M, \infty) \times (0, \pi) \times [0, 2 \pi)$, which we nonregularly extend into the interior region with $\textnormal{Ran}(r) = (0, 2 M)$. In this representation, the Schwarzschild metric is singular and causal-type-reversing at the event horizon at $r = 2 M$,\footnote{More precisely, this means that the gradient $\boldsymbol{\nabla}t$ associated with the Schwarzschild time coordinate $t$, which is timelike outside the event horizon, becomes spacelike inside the event horizon, and vice versa for the gradient $\boldsymbol{\nabla}r$ that corresponds to the Schwarzschild radial coordinate $r$ (see also Remark \ref{SchTime}).} so every light cone approaching the event horizon from the outside closes up, and, eventually, at the horizon degenerates,\footnote{This can be easily seen from the fact that at $r = 2 M$, $\textnormal{d}t/\textnormal{d}r \rightarrow \pm \infty$ along outgoing and ingoing radial null geodesics, respectively, causing the apertures of light cones to tend to zero.} whereas inside the horizon, light cones are tilted over such that all future-directed paths are pointing, instead of in the direction of increasing Schwarzschild time $t$ as in the exterior region, in the direction of decreasing Schwarzschild radius $r$. In order to remove these pathologies such that Conditions ($\mathcal{C}1$)--($\mathcal{C}3$) can be satisfied, we require a coordinate system that brings about a single, globally well-defined, possibly smooth representation of the metric covering the entire extension. A representation of this kind is naturally regular at the event horizon, where the horizon is to be found at finite coordinate values and not shifted to infinity, and furthermore preserves the causal types of the coordinates throughout the extension. To determine the desired coordinate system, we apply, in a first step, the transformation from Schwarzschild coordinates into Eddington--Finkelstein double-null coordinates\footnote{For a detailed discussion of the original Eddington--Finkelstein coordinates see Sections \ref{SubsubsectionVA} and \ref{SubsubsectionVB}. Their null variant, which we termed \textit{Penrose coordinates}, are reviewed in Section \ref{SubsubsectionVC}.}$^{,}$\footnote{A coordinate, say $u$, is called a null coordinate if the associated gradient $\boldsymbol{\nabla}u$ satisfies the null condition $\boldsymbol{g}(\boldsymbol{\nabla}u, \boldsymbol{\nabla}u) = 0$.}
\begin{equation*} 
\mathfrak{T}^{\textnormal{EF}} \colon
\begin{cases}
\, \mathbb{R} \times \mathbb{R}_{> 0} \times (0, \pi) \times [0, 2 \pi) \rightarrow \mathbb{R} \times \mathbb{R} \times (0, \pi) \times [0, 2 \pi) \\[0.25cm]
\hspace{2.57cm} (t, r, \vartheta, \phi) \mapsto (u, v, \vartheta', \phi') 
\end{cases} 
\end{equation*}
with 
\begin{equation} \label{DNC1} 
\left\{\!\begin{aligned}
& \, u = + t - r_{\star} & & \textnormal{and} \quad v = + t + r_{\star} & & \, \textnormal{for} \,\,\, \textnormal{B}_{\textnormal{I}} \\[0.1cm] 
& \, u = + t + r_{\star} & & \textnormal{and} \quad v = - t + r_{\star} & & \, \textnormal{for} \,\,\, \textnormal{B}_{\textnormal{II}} \\[0.1cm]
& \, u = - t + r_{\star} & & \textnormal{and} \quad v = - t - r_{\star} & & \, \textnormal{for} \,\,\, \textnormal{B}_{\textnormal{III}} \\[0.1cm] 
& \, u = - t - r_{\star} & & \textnormal{and} \quad v = + t - r_{\star} & & \, \textnormal{for} \,\,\, \textnormal{B}_{\textnormal{IV}}
\end{aligned}\right\} \, , \quad \vartheta' = \vartheta \, , \quad \textnormal{and} \quad \phi' = \phi \, ,
\end{equation}
where
\begin{equation} \label{RWC} 
r_{\star} := r + 2 M \, \ln{\bigg|\frac{r}{2 M} - 1\bigg|} \in \begin{cases}
\, \mathbb{R} & \, \textnormal{for} \,\,\,\,\, \textnormal{B}_{\textnormal{I}} \,\,\, \textnormal{and} \,\,\, \textnormal{B}_{\textnormal{III}} \\[0.1cm] 
\, \mathbb{R}_{< 0} & \, \textnormal{for} \,\,\, \textnormal{B}_{\textnormal{II}} \,\,\, \textnormal{and} \,\,\, \textnormal{B}_{\textnormal{IV}}
\end{cases}
\end{equation}
is the Regge--Wheeler coordinate, and $u < - v$ in $\textnormal{B}_{\textnormal{II}}$ and $u > - v$ in $\textnormal{B}_{\textnormal{IV}}$.\footnote{As in the case of Kruskal's coordinates for the maximal analytic extension presented in Section \ref{SubsectionIIB}, the particular forms of the transformation laws for the Eddington--Finkelstein double-null coordinates $u$ and $v$ in the regions $\textnormal{B}_{\textnormal{III}}$ and $\textnormal{B}_{\textnormal{IV}}$ follow directly from applying the discrete reversal isometries $u \mapsto u' = - u$ and $v \mapsto v' = - v$ to the transformation laws for the regions $\textnormal{B}_{\textnormal{I}}$ and $\textnormal{B}_{\textnormal{II}}$, respectively. Accordingly, the light cone structures in $\textnormal{B}_{\textnormal{III}}$ and $\textnormal{B}_{\textnormal{IV}}$ are reversed.} These coordinates are adapted to outgoing and ingoing radial null geodesics in the sense that the former geodesics satisfy the constraint $u = \textnormal{const.}$, whereas for the latter the constraint $v = \textnormal{const.}$ holds. Accordingly, if we decrease the Schwarzschild radial coordinate $r$ along a radial null curve with $u = \textnormal{const.}$ ($v = \textnormal{const.}$), we can now pass through the past (future) component of the event horizon both from region $\textnormal{B}_{\textnormal{I}}$ into region $\textnormal{B}_{\textnormal{IV}}$ ($\textnormal{B}_{\textnormal{II}}$) and from region $\textnormal{B}_{\textnormal{III}}$ into region $\textnormal{B}_{\textnormal{II}}$ ($\textnormal{B}_{\textnormal{IV}}$) in a regular manner and without changing the causal types of the coordinates [cf.\ the dashed lines in FIG.\ \ref{PenroseKruskalCoords}].\footnote{This is to say, the particular set of four different region-wise transformation laws specified in Equation (\ref{DNC1}) is arranged to obtain a consistent double-null coordinate covering of the maximal analytic extension $\medcup_{i = \textnormal{I}}^{\textnormal{IV}} \textnormal{B}_i$ of the exterior Schwarzschild spacetime that joins together the respective regions in a manner such that across abutting edges the coordinate curves of $u(t, r_{\star})$ and $v(t, r_{\star})$ are carried over smoothly, while properly accounting for the multivaluedness of the Schwarzschild time coordinate $t$ and of the Regge--Wheeler coordinate $r_{\star}$ [see the orientations of the $(u, v)$ and $(t, r_{\star})$ coordinate systems depicted in FIG.\ \ref{PenroseKruskalCoords}].}$^{,}$\footnote{In Schwarzschild coordinates, this was not possible for the Schwarzschild time coordinate $t$ along ingoing/outgoing radial null geodesics tends to $\pm \infty$ at the future/past components of the event horizon.} But although one may traverse the past and future components of the event horizon in the foregoing way, Eddington--Finkelstein double-null coordinates, similar to the Schwarzschild coordinates, cover each of the four regions of the maximal analytic extension only separately. Moreover, since the Schwarzschild metric expressed in Eddington--Finkelstein double-null coordinates reads
\begin{equation} \label{SMDNC}
\boldsymbol{g} = \frac{1}{2} \, \bigg|1 - \frac{2 M}{r(u, v)}\bigg| \, [\textnormal{d}u \otimes \textnormal{d}v + \textnormal{d}v \otimes \textnormal{d}u] - r(u, v)^2 \, \boldsymbol{g}_{S^2} 
\end{equation}
with the multivalued function\footnote{Here, the upper signs in the exponential functions correspond to the regions denoted to the left of the forward slash in the attributions, while the lower signs pertain to the regions to the right of the forward slash. We continue using this particular notation in the remainder of the paper whenever convenient.} 
\begin{equation*} 
r(u, v) = 2 M \left[1 + 
\left\{\!\begin{aligned}
& \, W_0\bigl(e^{\pm (v - u)/(4 M) - 1}\bigr) & & \, \textnormal{for} \,\,\, \textnormal{B}_{\textnormal{I}} \slash \textnormal{B}_{\textnormal{III}} \\[0.1cm] 
& \, W_0\bigl(- e^{\pm (v + u)/(4 M) - 1}\bigr) & & \, \textnormal{for} \,\,\, \textnormal{B}_{\textnormal{II}} \slash \textnormal{B}_{\textnormal{IV}}
\end{aligned} \right\}
\right]
\end{equation*}
[cf.\ the solution to Equation (\ref{ruv-eq}) given directly below and Remark \ref{SchwInt}], the metric determinant in this representation vanishes at $r = 2 M$, rendering the Schwarzschild metric not invertible at the event horizon, thus leaving it still degenerate.\footnote{In contrast, the determinant of the Schwarzschild representation of the nonregularly extended Schwarzschild metric is nonzero also at the event horizon at $r = 2 M$. This aspect does not, however, allow for any conclusions to be drawn about the invertibility of the metric at the event horizon, for the horizon is not in the range of the Schwarzschild radial coordinate, so that, strictly speaking, the metric determinant cannot be evaluated at the event horizon in the first place.} In addition, the event horizon has been pushed to infinity, viz., to $\bigl\{(u, v, \vartheta', \phi') \, \big| \, u \in \mathbb{R}, \, v = - \infty, \, \vartheta' \in (0, \pi), \, \phi' \in [0, 2 \pi)\bigr\}$ in $\textnormal{B}_{\textnormal{I}}$ and $\textnormal{B}_{\textnormal{II}}$, $\bigl\{(u, v, \vartheta', \phi') \, \big| \, u = \infty, \, v \in \mathbb{R}, \, \vartheta' \in (0, \pi), \, \phi' \in [0, 2 \pi)\bigr\}$ in $\textnormal{B}_{\textnormal{I}}$ and $\textnormal{B}_{\textnormal{IV}}$, $\bigl\{(u, v, \vartheta', \phi') \, \big| \, u = - \infty, \, v \in \mathbb{R}, \, \vartheta' \in (0, \pi), \, \phi' \in [0, 2 \pi)\bigr\}$ in $\textnormal{B}_{\textnormal{II}}$ and $\textnormal{B}_{\textnormal{III}}$, and $\bigl\{(u, v, \vartheta', \phi') \, \big| \, u \in \mathbb{R}, \, v = \infty, \, \vartheta' \in (0, \pi), \, \phi' \in [0, 2 \pi)\bigr\}$ in $\textnormal{B}_{\textnormal{III}}$ and $\textnormal{B}_{\textnormal{IV}}$. 
\begin{Remark} \label{AltCoords}
Working with coordinates $(t, r_{\star}, \vartheta, \phi) \in \mathbb{R} \times \mathbb{R} \times (0, \pi) \times [0, 2 \pi)$ for which the representation of the Schwarzschild metric becomes 
\begin{equation*}
\boldsymbol{g} = \biggl(1 - \frac{2 M}{r(r_{\star})}\biggr) \, [\textnormal{d}t \otimes \textnormal{d}t - \textnormal{d}r_{\star} \otimes \textnormal{d}r_{\star}] - r(r_{\star})^2 \, \boldsymbol{g}_{S^2} 
\end{equation*}
with the multivalued function 
\begin{equation*} 
r(r_{\star}) = 2 M \bigl[1 + W_0\bigl(\pm e^{r_{\star}/(2 M) - 1}\bigr)\bigr] \,\,\,\, \textnormal{for} \,\,\, \textnormal{B}_{\textnormal{I}} \,\,\, \textnormal{and} \,\,\, \textnormal{B}_{\textnormal{III}} \slash \textnormal{B}_{\textnormal{II}} \,\,\, \textnormal{and} \,\,\, \textnormal{B}_{\textnormal{IV}} \, ,
\end{equation*}
we encounter, aside from not being able to cross the event horizon in a regular, causal-type-preserving manner, analogous problems as for the above Eddington--Finkelstein double null coordinates, i.e., even though towards the event horizon light cones do not close up anymore,\footnote{Since $\textnormal{d}t/\textnormal{d}r_{\star} \rightarrow \pm 1$ along outgoing and ingoing radial null geodesics, respectively, the apertures of light cones expressed in these coordinates are $45^{\circ}$, being equal to the apertures of light cones in Minkowski spacetime.} the coordinates are only separately defined on the different regions of the maximal analytic extension, the metric representation is not invertible, and the location of the event horizon has been pushed to $r_{\star} = - \infty$. For advanced Eddington--Finkelstein coordinates $(v, r, \vartheta, \phi) \in \mathbb{R} \times \mathbb{R}_{> 0} \times (0, \pi) \times [0, 2 \pi)$, for example, the situation is different, as in this case the Schwarzschild time coordinate $t$ is replaced by a time coordinate $v = t + r_{\star} - r$ with a constant rate of change along ingoing radial null geodesics, so that progressions in the direction of the event horizon do not become continuously slower anymore.\footnote{The same holds for retarded Eddington--Finkelstein coordinates $(u, r, \vartheta, \phi) \in \mathbb{R} \times \mathbb{R}_{> 0} \times (0, \pi) \times [0, 2 \pi)$, where $u = t - r_{\star} + r$, except for the time coordinate $u$ having a constant rate of change along \textit{outgoing} radial null geodesics. For the advanced/retarded Eddington--Finkelstein null variants $(v/u, r, \vartheta, \phi) \in \mathbb{R} \times \mathbb{R}_{> 0} \times (0, \pi) \times [0, 2 \pi)$, the rate of change of the null coordinates $v/u = t \pm r_{\star}$ is zero, rendering them constants along ingoing/outgoing radial null geodesics.}$^{,}$\footnote{In contrast to Schwarzschild coordinates, where the time coordinate $t$ and the radial coordinate $r$ interchange their causal types across the event horizon, here only the causal type of the Eddington--Finkelstein radial coordinate $r$ reverses across the event horizon, i.e., the gradient $\boldsymbol{\nabla} r$ transitions from being spacelike in the exterior region to being timelike in the interior region. The causal type of the Eddington--Finkelstein time coordinate $v$, on the other hand, is timelike in both the exterior and the interior region. The latter observation is consistent with $v$ being a temporal function in $\textnormal{B}_{\textnormal{I}} \cup \textnormal{B}_{\textnormal{II}}$ (and similar for $\textnormal{B}_{\textnormal{III}} \cup \textnormal{B}_{\textnormal{IV}}$), that is to say that $v \colon \textnormal{B}_{\textnormal{I}} \cup \textnormal{B}_{\textnormal{II}} \rightarrow \mathbb{R}$ is a smooth function with future-directed, timelike gradient $\boldsymbol{\nabla}v$ whose level sets $v^{- 1}(\, . \,)$ are smooth, spacelike hypersurfaces (cf.\ Remark \ref{SchTime} for the stronger notion of Cauchy temporal functions).\label{fnvtemp}} Then, light cones remain well-behaved throughout the regions $\textnormal{B}_{\textnormal{I}} \cup \textnormal{B}_{\textnormal{II}}$ or $\textnormal{B}_{\textnormal{III}} \cup \textnormal{B}_{\textnormal{IV}}$,\footnote{To be more precise, approaching the event horizon from the outside, the outgoing component of every light cone continuously bends inwards until it coincides with the horizon, while the ingoing component remains unchanged. Inside the event horizon, the outgoing component still bends further inwards, but light cones never fully close up. As a consequence, all future-directed causal curves traversing the event horizon become trapped in the black hole region, ultimately impinging on the curvature singularity at $r = 0$.} and, since the advanced Eddington--Finkelstein representation of the Schwarzschild metric 
\begin{equation*}
\boldsymbol{g} = \biggl(1 - \frac{2 M}{r}\biggr) \, \textnormal{d}v \otimes \textnormal{d}v - \frac{2 M}{r} \, [\textnormal{d}v \otimes \textnormal{d}r + \textnormal{d}r \otimes \textnormal{d}v] - \biggl(1 + \frac{2 M}{r}\biggr) \, \textnormal{d}r \otimes \textnormal{d}r - r^2 \, \boldsymbol{g}_{S^2} 
\end{equation*}
gives rise to a nonvanishing determinant $\textnormal{det}(\boldsymbol{g}) =  - r^4 \sin^2(\vartheta)$ at $r = 2 M$, the metric is invertible at (the future component of) the event horizon. Besides, the horizon is located at finite coordinate values. However, advanced Eddington--Finkelstein coordinates cannot be analytically extended to cover the entire maximal analytic extension $\medcup_{i = \textnormal{I}}^{\textnormal{IV}} \textnormal{B}_i$.
\end{Remark}

To remove the remaining inadequacies, viz., the nonregularity of the metric at the event horizon, the multivaluedness of the coordinate ranges throughout the maximal analytic extension, and an event horizon location at infinity, while at the same time compactifying the coordinate ranges such that Conditions ($\mathcal{C}1$) and ($\mathcal{C}2$) can be met, we apply, in a second step, the transformation from Eddington--Finkelstein double-null coordinates into compactified Kruskal--Szekeres double-null coordinates 
\begin{equation*} 
\mathfrak{T}^{\textnormal{KS1}} \colon
\begin{cases}
\, \displaystyle \mathbb{R} \times \mathbb{R} \times (0, \pi) \times [0, 2 \pi) \rightarrow \displaystyle \biggl(- \frac{\pi}{2}, \frac{\pi}{2}\biggr) \times \biggl(- \frac{\pi}{2}, \frac{\pi}{2}\biggr) \times (0, \pi) \times [0, 2 \pi) \\[0.45cm]
\hspace{2.08cm} (u, v, \vartheta, \phi) \mapsto (U, V, \vartheta', \phi')
\end{cases} 
\end{equation*}
with 
\begin{equation} \label{DNC2} 
\left\{\!\begin{aligned}
& \, \tan{(U)} = - e^{- u/(4 M)} & & \textnormal{and} \quad \tan{(V)} = + e^{v/(4 M)} & & \, \textnormal{for} \,\,\, \textnormal{B}_{\textnormal{I}} \\[0.1cm] 
& \, \tan{(U)} = + e^{u/(4 M)} & & \textnormal{and} \quad \tan{(V)} = + e^{v/(4 M)} & & \, \textnormal{for} \,\,\, \textnormal{B}_{\textnormal{II}} \\[0.1cm]
& \, \tan{(U)} = + e^{u/(4 M)} & & \textnormal{and} \quad \tan{(V)} = - e^{- v/(4 M)} & & \, \textnormal{for} \,\,\, \textnormal{B}_{\textnormal{III}} \\[0.1cm] 
& \, \tan{(U)} = - e^{- u/(4 M)} & & \textnormal{and} \quad \tan{(V)} = - e^{- v/(4 M)} & & \, \textnormal{for} \,\,\, \textnormal{B}_{\textnormal{IV}}
\end{aligned}\right\} \, , \quad \vartheta' = \vartheta \, , \quad \textnormal{and} \quad \phi' = \phi \, ,
\end{equation}
where 
\begin{equation*} 
\left\{\!\begin{aligned}
& \, U \in (- \pi/2, 0) & & \textnormal{and} \quad V \in (0, \pi/2) & & \, \textnormal{for} \,\,\, \textnormal{B}_{\textnormal{I}} \\[0.1cm] 
& \, U \in (0, \pi/2 - V) & & \textnormal{and} \quad V \in (0, \pi/2) & & \, \textnormal{for} \,\,\, \textnormal{B}_{\textnormal{II}} \\[0.1cm]
& \, U \in (0, \pi/2) & & \textnormal{and} \quad V \in (- \pi/2, 0) & & \, \textnormal{for} \,\,\, \textnormal{B}_{\textnormal{III}} \\[0.1cm] 
& \, U \in (- \pi/2 - V, 0) & & \textnormal{and} \quad V \in (- \pi/2, 0) & & \, \textnormal{for} \,\,\, \textnormal{B}_{\textnormal{IV}}
\end{aligned} \right\} \, .\footnote{This particular compactification in terms of inverse tangent functions is chosen for its simplicity and convenience. However, since there is, a priori, no preferred choice of the compactification mapping, there are naturally many other suitable homeomorphic alternatives that map the given coordinate domains onto finite ranges.}
\end{equation*}
These coordinates are globally regular with a single-valued, bounded range and overall causal-type-preserving, i.e., they cover in a well-defined manner the entire maximal analytic extension $\medcup_{i = \textnormal{I}}^{\textnormal{IV}} \textnormal{B}_i$ of the exterior Schwarzschild spacetime compactified into a finite region. Moreover, they locate the event horizon at finite coordinate values at $\bigl\{(U, V, \vartheta', \phi') \, \big| \, U = 0, \, V \in (-\pi/2, \pi/2), \, \vartheta' \in (0, \pi), \, \phi' \in [0, 2 \pi)\bigr\}$ and $\bigl\{(U, V, \vartheta', \phi') \, \big| \, U \in (-\pi/2, \pi/2), \, V = 0, \, \vartheta' \in (0, \pi), \, \phi' \in [0, 2 \pi)\bigr\}$ away from the boundary. Expressed in compactified Kruskal--Szekeres double-null coordinates, the Schwarzschild metric becomes
\begin{equation*}
\boldsymbol{g} = \frac{16 M^3 \, e^{- r(U, V)/(2 M)}}{\cos^2(U) \cos^2(V) \, r(U, V)} \, [\textnormal{d}U \otimes \textnormal{d}V + \textnormal{d}V \otimes \textnormal{d}U] - r(U, V)^2 \, \boldsymbol{g}_{S^2}  
\end{equation*}
with the multivalued function 
\begin{equation*} 
r(U, V) = 2 M \bigl[1 + W_0\bigl(- e^{- 1} \tan(U) \tan(V)\bigr)\bigr] \quad \textnormal{for} \,\,\, \medcup_{i = \textnormal{I}}^{\textnormal{IV}} \textnormal{B}_i \, .
\end{equation*}
Represented in this form, the Schwarzschild metric is regular on the entire maximal analytic extension, making it is invertible for all values $r \in \mathbb{R}_{> 0}$ of the Schwarzschild radial coordinate. Besides, the $2$-dimensional Lorentzian component $\boldsymbol{g}_{\textnormal{L}}^{(2)}$ of this metric representation [cf.\ Equation (\ref{prodstr})] is now given in a form that is conformally equivalent to the $2$-dimensional Minkowski metric $\boldsymbol{\eta}^{(2)} = (\textnormal{d}u \otimes \textnormal{d}v + \textnormal{d}v \otimes \textnormal{d}u)/2$ described in double-null coordinates $\mathbb{R} \times \mathbb{R} \ni (u, v) = (t - r, t + r)$.\footnote{The Lorentzian component of the metric representation in Eddington--Finkelstein double-null coordinates specified in Equation (\ref{SMDNC}) is not already of conformally Minkowskian double-null form for the conformal factor, although being in the class of smooth functions on the maximal analytic extension, is not \textit{strictly} positive, viz., it vanishes at the event horizon at $r = 2 M$.}

Finally, for convenience and because spacetime coordinates are more customary than double-null coordinates, we transform the compactified Kruskal--Szekeres double-null coordinates into their spacetime variant.\footnote{Writing the Schwarzschild metric in terms of compactified Kruskal--Szekeres spacetime coordinates instead of compactified Kruskal--Szekeres double null coordinates not only makes the presence of the metrical product structure given in Equation (\ref{prodstr}) more visible, but it also directly shows its $2$-dimensional Lorentzian component $\boldsymbol{g}_{\textnormal{L}}^{(2)}$ to be conformally Minkowskian.} To this end, we employ the transformation
\begin{equation*} 
\mathfrak{T}^{\textnormal{KS2}} \colon
\begin{cases}
\, \displaystyle \biggl(- \frac{\pi}{2}, \frac{\pi}{2}\biggr) \times \biggl(- \frac{\pi}{2}, \frac{\pi}{2}\biggr) \times (0, \pi) \times [0, 2 \pi)  \rightarrow \displaystyle \biggl(- \frac{\pi}{4}, \frac{\pi}{4}\biggr) \times \biggl(- \frac{\pi}{2}, \frac{\pi}{2}\biggr) \times (0, \pi) \times [0, 2 \pi) \\[0.45cm]
\hspace{4.58cm} (U, V, \vartheta, \phi) \mapsto (T, X, \vartheta', \phi') 
\end{cases} 
\end{equation*}
with 
\begin{equation} \label{TXUV} 
T = \frac{U + V}{2} \, , \quad X = \frac{- U + V}{2} \, , \quad \vartheta' = \vartheta \, , \quad \textnormal{and} \quad \phi' = \phi \quad \textnormal{for} \,\, \medcup_{i = \textnormal{I}}^{\textnormal{IV}} \textnormal{B}_i \, ,
\end{equation}
where 
\begin{equation} \label{TXranges}
\left\{\!\begin{aligned}
& \, T \in \bigl( \hspace{0.02cm} |X - \pi/4| - \pi/4, - |X - \pi/4| + \pi/4\bigr) & & \textnormal{and} \quad X \in (0, \pi/2) & & \, \textnormal{for} \,\,\, \textnormal{B}_{\textnormal{I}} \\[0.1cm] 
& \, T \in \bigl( \hspace{0.02cm} |X|, \pi/4\bigr) & & \textnormal{and} \quad X \in (- \pi/4, \pi/4) & & \, \textnormal{for} \,\,\, \textnormal{B}_{\textnormal{II}} \\[0.1cm]
& \, T \in \bigl( \hspace{0.02cm} |X + \pi/4| - \pi/4, - |X + \pi/4| + \pi/4\bigr) & & \textnormal{and} \quad X \in (- \pi/2, 0) & & \, \textnormal{for} \,\,\, \textnormal{B}_{\textnormal{III}} \\[0.1cm] 
& \, T \in \bigl(- \pi/4, - |X| \hspace{0.02cm} \bigr) & & \textnormal{and} \quad X \in (- \pi/4, \pi/4) & & \, \textnormal{for} \,\,\, \textnormal{B}_{\textnormal{IV}}
\end{aligned} \right\} \, .\footnote{Geometrically, this particular coordinate transformation constitutes a combined rotation of the $U$-$V$ plane about $45^{\circ}$ counterclockwise and scaling of the $U$ and $V$ axes by a factor of $1/\sqrt{2}$. Choosing a normalization constant of value $1/\sqrt{2}$ instead of $1/2$ in the definition of the coordinates $T$ and $X$ would render the transformation just a simple SO$(2)$ rotation of the above type, but cause more inconvenient coordinate ranges.}^{,}\footnote{Expressed in terms of the Schwarzschild time coordinate $t$ and the Regge--Wheeler coordinate $r_{\star}$ defined in Equation (\ref{RWC}), the compactified Kruskal--Szekeres time and radial coordinates yield 
\begin{equation*} 
\left\{\!\begin{aligned}
& \, T = \pm \frac{1}{2} \arctan{\Biggl(\frac{\sinh{\bigl(t/(4 M)\bigr)}}{\cosh{\bigl(r_{\star}/(4 M)\bigr)}}\Biggr)} \quad \textnormal{and} \quad X = \mp \frac{1}{2} \arctan{\Biggl(\frac{\cosh{\bigl(t/(4 M)\bigr)}}{\sinh{\bigl(r_{\star}/(4 M)\bigr)}}\Biggr)} \pm \frac{\pi H(r_{\star})}{2} & & \, \textnormal{for} \,\,\, \textnormal{B}_{\textnormal{I}}\slash \textnormal{B}_{\textnormal{III}} \\[0.1cm] 
& \, T = \mp \frac{1}{2} \arctan{\Biggl(\frac{\cosh{\bigl(t/(4 M)\bigr)}}{\sinh{\bigl(r_{\star}/(4 M)\bigr)}}\Biggr)} \quad \hspace{0.03cm} \textnormal{and} \quad X = \mp \frac{1}{2} \arctan{\Biggl(\frac{\sinh{\bigl(t/(4 M)\bigr)}}{\cosh{\bigl(r_{\star}/(4 M)\bigr)}}\Biggr)} & & \, \textnormal{for} \,\,\, \textnormal{B}_{\textnormal{II}}\slash \textnormal{B}_{\textnormal{IV}} 
\end{aligned} \right\} \, ,
\end{equation*}
where $H(\, . \,) := [1 + \textnormal{sgn}(\, . \, )]/2$ is the usual Heaviside step function.}^{,}\footnote{Similar to the Schwarzschild time coordinate $t$ in the region $\textnormal{B}_{\textnormal{I}}$ or the region $\textnormal{B}_{\textnormal{III}}$, the compactified Kruskal--Szekeres time coordinate $T$ is a Cauchy temporal function in the maximal analytic extension $\medcup_{i = \textnormal{I}}^{\textnormal{IV}} \textnormal{B}_i$ of the exterior Schwarzschild spacetime (cf.\ Remark \ref{SchTime}).}
\end{equation}
The Schwarzschild metric represented in these coordinates reads
\begin{equation} \label{KSSTC} 
\boldsymbol{g} = \frac{32 M^3 \, e^{- r(T, X)/(2 M)}}{\bigl[\cos^2(T) - \sin^2(X)\bigr]^2 \, r(T, X)} \, [\textnormal{d}T \otimes \textnormal{d}T - \textnormal{d}X \otimes \textnormal{d}X] - r(T, X)^2 \, \boldsymbol{g}_{S^2} \, , 
\end{equation}
with the multivalued function $r(T, X)$ also being determined by the principal branch of the Lambert $W$ function, 
\begin{equation*} 
r(T, X) = 2 M \bigl[1 + W_0\bigl(- e^{- 1} \tan(T + X) \tan(T - X)\bigr)\bigr] \quad \textnormal{for} \,\,\, \medcup_{i = \textnormal{I}}^{\textnormal{IV}} \textnormal{B}_i \, .
\end{equation*}

\subsection{Components of the Penrose Diagram} \label{SubsectionIVB}

\noindent To explicitly construct and draw the Penrose diagram of the maximally analytically extended Schwarzschild spacetime, we have to identify the diagram's various components, viz., its boundary components and its interior regions, utilizing the compactified Kruskal--Szekeres time and radial coordinates $T$ and $X$ defined in Equation (\ref{TXUV}) with ranges specified in Equation (\ref{TXranges}). To this end, it is convenient to employ the transcendental algebraic equations 
\begin{equation} \label{TAERWC} 
\left\{\!\begin{aligned}
& \, \displaystyle \frac{\cos{(2 T)}}{\cos{(2 X)}} = - \coth{\bigl(r_{\star}(r)/(4 M)\bigr)} \, \hspace{0.37cm} \textnormal{for} \,\,\, \textnormal{B}_{\textnormal{I}} \,\,\, \textnormal{and} \,\,\, \textnormal{B}_{\textnormal{III}} \\[0.15cm] 
& \, \displaystyle \frac{\cos{(2 T)}}{\cos{(2 X)}} = - \tanh{\bigl(r_{\star}(r)/(4 M)\bigr)} \, \hspace{0.3cm} \textnormal{for} \,\, \textnormal{B}_{\textnormal{II}} \,\,\, \textnormal{and} \,\,\, \textnormal{B}_{\textnormal{IV}} 
\end{aligned}\right\} 
\end{equation}
that relate the compactified Kruskal--Szekeres time and radial coordinates to the Regge--Wheeler coordinate and those relating the former to the Schwarzschild time coordinate
\begin{equation} \label{TAERWC2}
\left\{\!\begin{aligned}
& \, \displaystyle \frac{\sin{(2 T)}}{\sin{(2 X)}} = + \tanh{\bigl(t/(4 M)\bigr)} \, \hspace{0.34cm} \textnormal{for} \,\,\, \textnormal{B}_{\textnormal{I}} \,\,\, \textnormal{and} \,\,\,  \textnormal{B}_{\textnormal{III}} \\[0.15cm] 
& \, \displaystyle \frac{\sin{(2 T)}}{\sin{(2 X)}} = - \coth{\bigl(t/(4 M)\bigr)} \, \hspace{0.35cm} \textnormal{for} \,\, \textnormal{B}_{\textnormal{II}} \,\,\, \textnormal{and} \,\,\, \textnormal{B}_{\textnormal{IV}} 
\end{aligned}\right\} ,
\end{equation}
from which one can directly read off the asymptotic relations presented in TABLE I.\footnote{The transcendental algebraic equations defined in Equations (\ref{TAERWC}) and (\ref{TAERWC2}) can be straightforwardly derived through the transformations laws specified in Equations (\ref{DNC1}), (\ref{DNC2}), and (\ref{TXUV}) and basic trigonometric identities. More precisely, exemplified for the first transcendental algebraic equation given in Equation (\ref{TAERWC}) for the region $\textnormal{B}_{\textnormal{I}}$, we apply, in a first step, the transformation laws in Equation (\ref{TXUV}), which relate the compactified Kruskal--Szekeres time and radial coordinates $T$ and $X$ to the compactified Kruskal--Szekeres double-null coordinates $U$ and $V$, to the expression $\cos(2 T)/\cos(2 X)$. In a second step, we use angle addition theorems, the Pythagorean trigonometric identity, and simple algebraic manipulations to obtain the relation
\begin{equation} \label{TrigID}
\frac{\cos(2 T)}{\cos(2 X)} = \frac{1 - \tan(U) \tan(V)}{1 + \tan(U) \tan(V)} \, . 
\end{equation}
Subsequently, in a third step, inserting the transformation laws for the Eddington--Finkelstein double-null coordinates $u$ and $v$ defined in Equation (\ref{DNC1}) for the region $\textnormal{B}_{\textnormal{I}}$ into the corresponding transformation laws in Equation (\ref{DNC2}), and eliminate the functional dependence on the Schwarzschild time coordinate $t$, we find the relation
\begin{equation*}
- \tan(U) \tan(V) = e^{r_{\star}/(2 M)} 
\end{equation*}
between the compactified Kruskal--Szekeres double-null coordinates and the Regge--Wheeler coordinate $r_{\star}$. Substitution into Equation (\ref{TrigID}) immediately yields the desired result. We note that the particular shapes of the transcendental algebraic equations were chosen for their simplicity and convenience in the evaluation of the asymptotic relations between the Schwarzschild and Kruskal--Szekeres time and radial coordinates shown in TABLE I.} These asymptotics can then be employed to define the necessary exterior and interior boundary components of the Penrose diagram of the maximally analytically extended Schwarzschild spacetime:
\begin{itemize}
\item[(a)] Future/past timelike infinity $i^{\pm} = (T = \pm \pi/4, X = \pi/4 \,\,\, \textnormal{in} \,\,\, \textnormal{B}_{\textnormal{I}} \,\,\, \textnormal{and} \,\,\,  X = - \pi/4 \,\,\, \textnormal{in} \,\,\, \textnormal{B}_{\textnormal{III}})$.
\item[(b)] Future/past null infinity $\mathscr{I}^{\pm} = \bigl\{(T, X) \, \big| \, T = \pm (- X + \pi/2) \,\,\, \textnormal{and} \,\,\, \pi/4 < X < \pi/2 \,\,\, \textnormal{in} \,\,\, \textnormal{B}_{\textnormal{I}} \,\,\, \textnormal{and} \,\,\,  T = \pm (X + \pi/2) \,\,\, \textnormal{and} \, - \pi/2 < X < - \pi/4 \,\,\, \textnormal{in} \,\,\, \textnormal{B}_{\textnormal{III}}\bigr\}$.
\item[(c)] Spacelike infinity $i^0 = (T = 0, X = \pm \pi/2)$.
\item[(d)] The curvature singularity at $\bigl\{(T, X) \, \big| \, T = \pm \pi/4 \,\,\,  \textnormal{and} \, - \pi/4 < X < \pi/4\bigr\}$.
\item[(e)] The event horizon $\mathfrak{H} = \mathfrak{H}^+ \cup \mathfrak{H}^-$ with future/past components $\mathfrak{H}^{\pm} = \bigl\{(T, X) \, \big| \, T = \pm X \,\,\, \textnormal{and} \, - \pi/4 \leq X \leq \pi/4\bigr\}$.
\end{itemize}
\begin{table}[t] \label{table1}
\caption{Asymptotic relations between the Schwarzschild and Kruskal--Szekeres time and radial coordinates.}
\begin{ruledtabular}
\begin{tabular}{llll}
\\[-0.2cm]
& \hspace{1.9cm} $r \rightarrow \infty$ & \hspace{1.7cm} $r \rightarrow 2 M$ & \hspace{1.3cm} $r \rightarrow 0$ \\ \\
$\textnormal{B}_{\textnormal{I}}$ & $T = \pm \, [X - \pi/2], X \in [\pi/4, \pi/2]$ & \hspace{0.5cm} $T = \pm X, X \in [0, \pi/4]$ & \hspace{1.55cm} --- \\ \\
$\textnormal{B}_{\textnormal{II}}$ & \hspace{2.2cm} --- & \hspace{0.5cm} $T = |X|, X \in [- \pi/4, \pi/4]$ & $T = \pi/4, X \in (- \pi/4, \pi/4)$ \\ \\
$\textnormal{B}_{\textnormal{III}}$ & $T = \pm \, [X + \pi/2], X \in [- \pi/2, - \pi/4]$ & \hspace{0.5cm} $T = \pm X, X \in [- \pi/4, 0]$ & \hspace{1.55cm} --- \\ \\
$\textnormal{B}_{\textnormal{IV}}$ & \hspace{2.2cm} --- & \hspace{0.5cm} $T = - |X|, X \in [- \pi/4, \pi/4]$ & $T = - \pi/4, X \in (- \pi/4, \pi/4)$ \\[0.2cm]
\hline \\[-0.2cm]
& \hspace{1.9cm} $t \rightarrow \infty$ & \hspace{1.6cm} $t \rightarrow - \infty$ & \\ \\
$\textnormal{B}_{\textnormal{I}}$ & $T = - [X - \pi/2], X \in [\pi/4, \pi/2]$ & \hspace{0.5cm} $T = [X - \pi/2], X \in [\pi/4, \pi/2]$ & \\ \\
& $T = X, X \in [0, \pi/4]$ & \hspace{0.5cm} $T = - X, X \in [0, \pi/4]$ & \\ \\
$\textnormal{B}_{\textnormal{II}}$ & $T = - X, X \in [- \pi/4, 0]$ & \hspace{0.5cm} $T = X, X \in [0, \pi/4]$ & \\ \\
$\textnormal{B}_{\textnormal{III}}$ & $T = - [X + \pi/2], X \in [- \pi/2, - \pi/4]$ & \hspace{0.5cm} $T = [X + \pi/2], X \in [- \pi/2, - \pi/4]$ & \\ \\
& $T = X, X \in [- \pi/4, 0]$ & \hspace{0.5cm} $T = - X, X \in [- \pi/4, 0]$ & \\ \\
$\textnormal{B}_{\textnormal{IV}}$ & $T = - X, X \in [0, \pi/4]$ & \hspace{0.5cm} $T = X, X \in [- \pi/4, 0]$ & \\[0.2cm]
\end{tabular}
\end{ruledtabular}
\end{table}

\begin{figure}[t]%
\centering
\subfigure[][]{%
\label{PenroseKruskal}%
\includegraphics[width=0.48\columnwidth]{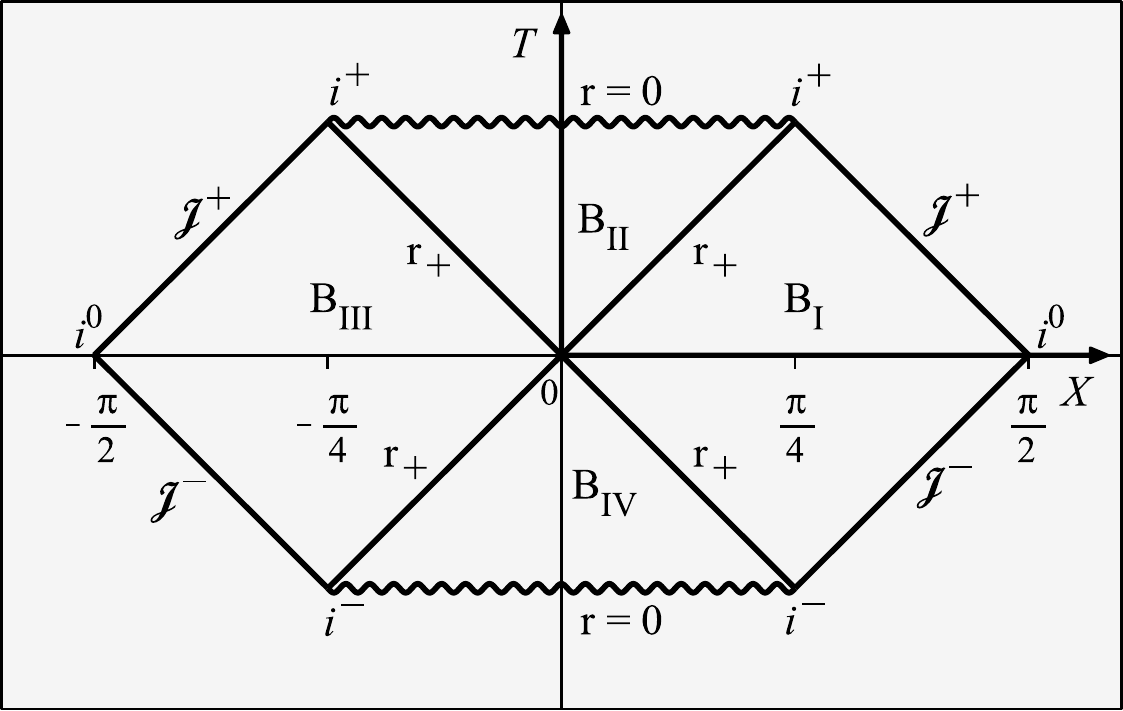}}
\hspace{8pt}%
\subfigure[][]{%
\label{PenroseKruskalCoords}%
\includegraphics[width=0.48\columnwidth]{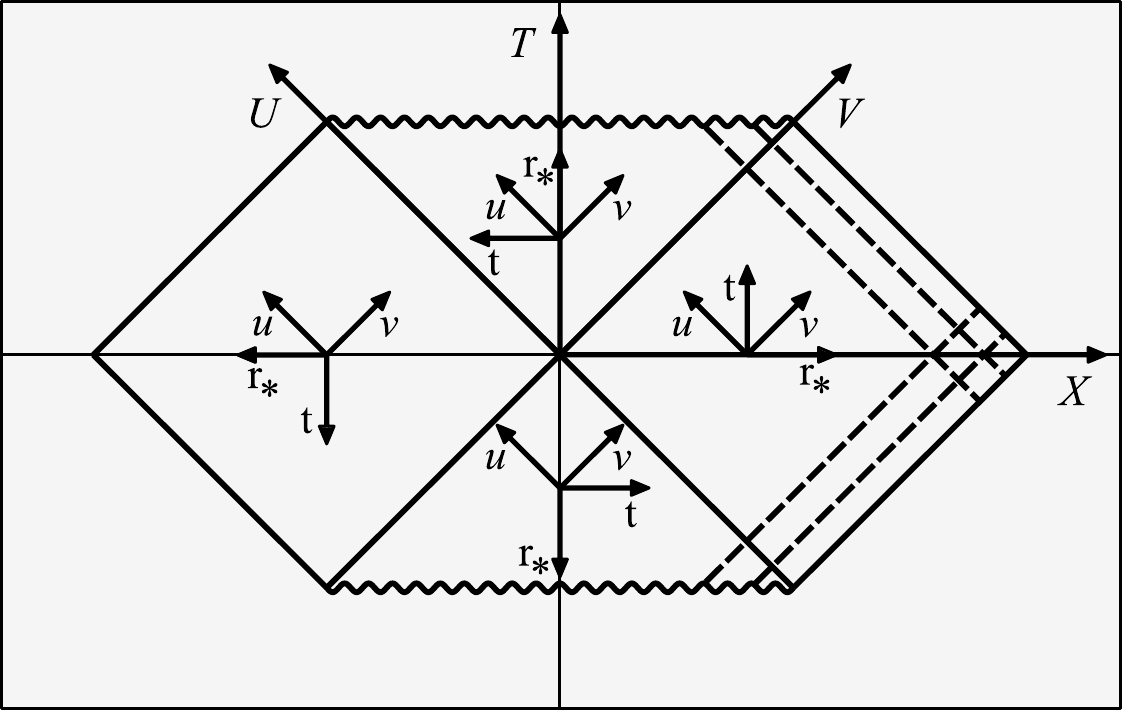}} 
\caption[...]
{\subref{PenroseKruskal} Penrose diagram of the maximally analytically extended Schwarzschild spacetime with labels of the exterior and interior boundary components and the different interior regions. \subref{PenroseKruskalCoords} The same diagram showing the orientations of the coordinate systems paramount for its construction (scalings are suppressed for simplicity) and depicting particular null hypersurfaces with either $u = \textnormal{const.}$ or $v = \textnormal{const.}$ indicated by the dashed diagonal lines.}%
\label{PenroseKruskalSlice}%
\end{figure}

\noindent Accordingly, the total exterior boundary of the diagram comprises conformal infinity and the curvature singularity, whereas the inner boundary consists of the event horizon. The interior regions of the Penrose diagram, which are enclosed by the exterior boundary components and separated by the interior boundary components, thus consist of the following four connected regions (cf.\ the end of Section \ref{SubsectionIIB}):
\begin{equation*}
\begin{split}
\textnormal{B}_{\textnormal{I}}^{(2)} & = \bigl\{(T, X) \, \big| \, T \in \bigl(|X - \pi/4| - \pi/4, - |X - \pi/4| + \pi/4\bigr) \,\,\, \textnormal{and} \,\,\, X \in (0, \pi/2)\bigr\} \\[0.2cm]
\textnormal{B}_{\textnormal{II}}^{(2)} & = \bigl\{(T, X) \, \big| \, T \in \bigl(|X|, \pi/4\bigr) \,\,\, \textnormal{and} \,\,\, X \in (- \pi/4, \pi/4)\bigr\} \\[0.2cm] 
\textnormal{B}_{\textnormal{III}}^{(2)} & = \bigl\{(T, X) \, \big| \, T \in \bigl(|X + \pi/4| - \pi/4, - |X + \pi/4| + \pi/4\bigr) \,\,\, \textnormal{and} \,\,\, X \in (- \pi/2, 0)\bigr\} \\[0.2cm]
\textnormal{B}_{\textnormal{IV}}^{(2)} & = \bigl\{(T, X) \, \big| \, T \in \bigl(- \pi/4, - |X|\bigr) \,\,\, \textnormal{and} \,\,\, X \in (- \pi/4, \pi/4)\bigr\} \, .
\end{split}
\end{equation*}
\noindent These regions represent the exterior region, the black hole region, the parallel exterior region, and the white hole region, respectively.\footnote{In the remainder of the paper, we drop the superscript ``$(2)$'' in the designators for these $2$-dimensional regions for simplicity and convenience.} Together with the boundary components (a)--(e), they make up the Penrose diagram of the entire maximally analytically extended Schwarzschild spacetime depicted in FIG.\ \ref{PenroseKruskal}. The metric $\boldsymbol{g}_{\textnormal{L}}^{(2)}$ on this diagram is given by the metric in Equation (\ref{KSSTC}) with the angular component $\boldsymbol{g}_{S^2}$ projected out.

%
%
\begin{Remark}
The point $(T = 0, X = 0)$, the so-called \textit{bifurcation $2$-sphere}, is usually identified with a nontraversable wormhole, known as an Einstein--Rosen bridge, connecting the two separated exterior regions $\textnormal{B}_{\textnormal{I}}$ and $\textnormal{B}_{\textnormal{III}}$.
\end{Remark}
\begin{Remark} \label{BHRandEH}
The exterior boundary component future null infinity $\mathscr{I}^+$, which is a connected null curve with past limit point at spacelike infinity $i^0$, is canonically used in a rigorous definition of the black hole region $\textnormal{B}_{\textnormal{II}}$, namely $\mathfrak{M} \backslash J^-(\mathscr{I}^+) \not= \emptyset$. Being the past boundary of this region, the future component $\mathfrak{H}^+$ of the event horizon can be therefore specified as $\partial J^-(\mathscr{I}^+) \cap \mathfrak{M}$ (see Footnote 16). By interchanging the designators ``future'' and ``past,'' one may similarly define the white hole region $\textnormal{B}_{\textnormal{IV}}$ and the past component $\mathfrak{H}^-$ of the event horizon.
\end{Remark}
\begin{Remark}
Although one might get the impression that in the Penrose diagram shown in FIG.~\ref{PenroseKruskal} the curvature singularity at $r = 0$ is represented by curly spacelike curves that span from future/past timelike infinity $i^{\pm}$ in $\textnormal{B}_{\textnormal{I}}$ to the respective counterparts in $\textnormal{B}_{\textnormal{III}}$, the infinities are indeed distinct from the curvature singularity, and thus from the two curly spacelike curves. This can be directly seen from the simple fact that all future/past-directed, timelike curves in $\textnormal{B}_{\textnormal{I}}$ that do not cross the event horizon never encounter the curvature singularity but eventually reach future/past timelike infinity, whereas every future/past-directed, timelike curve in $\textnormal{B}_{\textnormal{II}}/\textnormal{B}_{\textnormal{IV}}$ reaches the curvature singularity in finite eigentime. 
\end{Remark}
\begin{Remark} \label{GCP}
The above construction procedure of the Penrose diagram of the maximally analytically extended Schwarzschild spacetime, which is essentially based on the $2$-dimensional Lorentzian component
\begin{equation*} 
\boldsymbol{g}_{\textnormal{L}}^{(2)} = \biggl(1 - \frac{2 M}{r}\biggr) \textnormal{d}t \otimes \textnormal{d}t - \biggl(1 - \frac{2 M}{r}\biggr)^{- 1} \, \textnormal{d}r \otimes \textnormal{d}r
\end{equation*}
of Schwarzschild's metric representation (\ref{SchwarzschildRep}), can be directly generalized to a construction procedure of Penrose diagrams of spacetimes with metrics having Lorentzian components of the form
\begin{equation} \label{GL2M}
\boldsymbol{g}_{\textnormal{L}}^{(2)} = \mathscr{F}(r) \, \mathscr{H}_1(r) \, \textnormal{d}t \otimes \textnormal{d}t - \mathscr{F}(r)^{- 1} \, \mathscr{H}_2(r) \, \textnormal{d}r \otimes \textnormal{d}r \, ,
\end{equation}
where $\mathscr{F} \in C^{\infty}(\mathbb{R}, \mathbb{R})$ is analytic and $\mathscr{H}_{1/2} \in C^{\infty}(\mathbb{R}, \mathbb{R}_{> 0})$ \cite{ChruscielÖlzSzybka}. The different steps of this generalized construction procedure are similar to the ones of the Schwarzschild case, i.e., one first introduces the generalized Regge--Wheeler coordinate 
\begin{equation} \label{GRWC} 
r_{\star}(r) = \int^r_c \widetilde{\mathscr{F}}(s)^{- 1} \, \textnormal{d}s \, ,
\end{equation}
where $\widetilde{\mathscr{F}} := \mathscr{F} \sqrt{\mathscr{H}_1/\mathscr{H}_2}$ and the value of the constant $c$ has to be chosen such that $\widetilde{\mathscr{F}}(c) \not= 0$ [cf.\ Equation (\ref{RWC})], and identifies disjoint, maximal intervals in $\mathbb{R}$ on which the function $\widetilde{\mathscr{F}}$ is both finite and does not change sign. This gives rise to connected Lorentzian $2$-manifolds, referred to as Boyer--Lindquist blocks, on which the $2$-dimensional Lorentzian component $\boldsymbol{g}_{\textnormal{L}}^{(2)}$ specified in Equation (\ref{GL2M}) is separately well-defined, and which are to be suitably patched together at their respective conjunctive null hypersurfaces determined by the condition $\widetilde{\mathscr{F}} = 0$.\footnote{In the Schwarzschild case, the disjoint maximal intervals are given by $(2 M, \infty)$ and $(0, 2 M)$. The conjunctive null hypersurface is the event horizon at $r = 2 M$.}$^{,}$\footnote{Although the particular value of the constant $c$ in the generalized Regge--Wheeler coordinate (\ref{GRWC}) could, a priori, be chosen differently in each block, a careful analysis shows it to be necessarily identical everywhere (for the details see \cite{SchindlerAguirre}).} Subsequently, one also defines a pair of multivalued Eddington--Finkelstein-type double-null coordinates $(u, v) \in \mathbb{R} \times \mathbb{R}$ with possibly infinite ranges on each block [cf.\ Equation (\ref{DNC1})], which is finally transformed into a globally regular, single-valued compactified Kruskal--Szekeres-type double-null coordinate system $(U, V) \in \mathcal{I}_1 \times \mathcal{I}_2$ with finite ranges $\mathcal{I}_i \subset \mathbb{R}$, $i \in \{1, 2\}$, covering the totality of all blocks [cf.\ Equation (\ref{DNC2})].
\end{Remark}
%
%

\section{Application to the Eddington--Finkelstein and Penrose Foliations of the Schwarzschild Spacetime} \label{SectionV}

\noindent In the following, we present the Penrose diagram of the maximally analytically extended Schwarzschild spacetime equipped with, on the one hand, a foliation by the level sets of the time coordinate of the canonical Eddington--Finkelstein representation and, on the other hand, a foliation by the level sets of the null coordinate of the related Penrose representation. These diagrams allow us to easily work out the main properties of---and hence the differences between---the two coordinate representations by direct visual inspection. To have the proper background, we recall the main results of the seminal research papers by Eddington \cite{Eddington}, Finkelstein \cite{Finkelstein}, and Penrose \cite{Penrose3} pertaining to their respective coordinate representation of the Schwarzschild spacetime.

\subsection{Historical Remarks on the Eddington--Finkelstein and Penrose Coordinate Representations} \label{SubsectionVA}

\subsubsection{Eddington Coordinates} \label{SubsubsectionVA}

\noindent We begin with the derivation of Eddington's coordinates for the exterior Schwarzschild solution detailed in his 1924 paper \cite{Eddington}, which, contrary to what one may expect, was not motivated by a resolution of the Schwarzschild singularity at $r = 2 M$, uncovering that it is a mere coordinate singularity. It was rather a technical byproduct in a short mathematical proof to demonstrate his claim of an exact equality of the Schwarzschild line element $\textnormal{d}s$ of Einstein's general theory of relativity and the corresponding quantity $\textnormal{d}J$ of Whitehead's theory of gravitation \cite{Whitehead}, both being integral to the determination of the paths taken by test particles in the gravitational field of a single massive point particle.\footnote{To substantiate his equality assumption, Eddington points to the fact that the observed perihelion precession of Mercury and the observed deflection of light by the Sun can be explained consistently within the frameworks of both theories, so ``$\textnormal{d}J$ cannot be widely different from $\textnormal{d}s$ in the field of a single particle (the sun).''}$^{,}$\footnote{A strict equality between the expressions for $\textnormal{d}s$ and $\textnormal{d}J$ does not, however, exist in the case of multiple gravitational-field-generating massive point particles. This is a consequence of the fact that in general relativity, the overall gravitational effect of these point particles arises from the theory-specific nonlinear coupling, whereas in Whitehead's theory, it results from a mere superposition of the point particles' separate gravitational effects.}

More precisely, in his proof, Eddington considers the gravitational field generated by a single massive point particle at rest, which he describes in terms of the squared Schwarzschild line element expressed in Schwarzschild coordinates\footnote{Instead of the usual symbol $t$, which Eddington reserves for Whitehead's time coordinate, he denotes the Schwarzschild time coordinate by $t_1 \in \mathbb{R}$.}
\begin{equation*} 
\textnormal{d}s^2 = \biggl(1 - \frac{2 M}{r}\biggr) \textnormal{d}t_1^2 - \biggl(1 - \frac{2 M}{r}\biggr)^{- 1} \textnormal{d}r^2 - r^2 \bigl[\textnormal{d}\vartheta^2 + \sin^2(\vartheta) \, \textnormal{d}\phi^2\bigr] \, .
\end{equation*}
In order to bring this squared line element into the particular form of Whitehead's expression for $\textnormal{d}J^2$, he uses the coordinate transformation from Schwarzschild coordinates to Whitehead's coordinates
\begin{equation*} 
\mathfrak{T}^{(\textnormal{E})} \colon
\begin{cases}
\, \mathbb{R} \times (2 M, \infty) \times (0, \pi) \times [0, 2 \pi) \rightarrow \mathbb{R} \times (2 M, \infty) \times (0, \pi) \times [0, 2 \pi)  \\[0.25cm]
\hspace{3.1cm} (t_1, r, \vartheta, \phi) \mapsto (t, r', \vartheta', \phi') 
\end{cases} 
\end{equation*}
with
\begin{equation*}
t = t_1 - 2 M \ln{(r - 2 M)} \, , \quad r' = r \, , \quad \vartheta' = \vartheta \, , \quad \textnormal{and} \quad \phi' = \phi \, .\footnote{Actually, Eddington employs the time coordinate $t = t_1 - 2 M \log{(r - M)}$, where $\log(\, . \,)$ is regarded as the natural logarithm and, instead of $r - 2 M$, the argument of the logarithm wrongly reads $r - M$ without the required prefactor of $2$ in front of the second term.}  
\end{equation*}
This coordinate transformation not only ensures that the (normalized) speed of light along the outward radial direction is unity everywhere, that being a paramount aspect of Whitehead's theory of gravitation, but it also gives rise to the desired representation of the squared Schwarzschild line element
\begin{equation*} 
\textnormal{d}s^2 = \textnormal{d}s^2_{\mathbb{R}^{1, 3}} - \frac{2 M}{r} \, [\textnormal{d}t - \textnormal{d}r]^2 \, ,
\end{equation*}
which is of exactly the same form as $\textnormal{d}J^2$ in Whitehead's theory.\footnote{Here, $\textnormal{d}s_{\mathbb{R}^{1, 3}}$ is the line element in $4$-dimensional Minkowski spacetime $\mathbb{R}^{1, 3}$.}$^{,}$\footnote{The associated metric in today's canonical standard form is given by
\begin{equation*} 
\boldsymbol{g} = \biggl(1 - \frac{2 M}{r}\biggr) \textnormal{d}t \otimes \textnormal{d}t - \biggl(1 + \frac{2 M}{r}\biggr) \textnormal{d}r \otimes \textnormal{d}r + \frac{2 M}{r} \, [\textnormal{d}t \otimes \textnormal{d}r + \textnormal{d}r \otimes \textnormal{d}t] - r^2 \boldsymbol{g}_{S^2} \, .
\end{equation*}}
\begin{Remark} \label{EddProps}
Eddington coordinates are proper spacetime coordinates with a temporal function $t \colon \textnormal{B}_{\textnormal{I}} \rightarrow \mathbb{R}$ (cf.\ Footnote \ref{fnvtemp}). Moreover, although they were originally only defined on the exterior region $\textnormal{B}_{\textnormal{I}}$, they may be analytically extended across the past component of the event horizon at $r = 2 M$ to also cover the white hole region $\textnormal{B}_{\textnormal{IV}}$ (using the modified Eddington time coordinate $t = t_1 - 2 M \ln{|r - 2 M|}$). Then, the Schwarzschild metric expressed in the Eddington representation is time-independent and regular throughout $\textnormal{B}_{\textnormal{I}} \cup \textnormal{B}_{\textnormal{IV}}$.   
\end{Remark}

\subsubsection{Finkelstein Coordinates} \label{SubsubsectionVB}

\noindent Next, we proceed with Finkelstein's seminal 1958 paper \cite{Finkelstein} on a potential past-future asymmetry of the gravitational field of a massive, spherically symmetric point particle, where he constructs what he calls the \textit{complete} analytic extension of the exterior Schwarzschild solution, presumed both to hold throughout all of the particle's (empty) spacetime and not to possess any kind of singularities\footnote{Finkelstein employs the broader term ``irregularities.''} other than the one at the location of the point particle itself. Based on this particular analytic extension, he intends to show that although the Einstein field equations are invariant under time reversals, i.e., they do not single out a preferred direction of time, there exist solutions to these equations, specifically the solution corresponding to the gravitational field of a massive, spherically symmetric point particle, for which such a distinction---and thus a past-future asymmetry---naturally arises.\footnote{More precisely, according to Finkelstein, the actual metric constituting the spacetime of a massive, spherically symmetric point particle turns out to be not invariant under time reversal transformations $t \mapsto \overline{t} = - t$ for \textit{any} admissible choice of time coordinate.}

In more detail, Finkelstein models the spacetime $(\mathfrak{M}, \boldsymbol{g})$ of an isolated, massive, spherically symmetric point particle, on the one hand, as the entirety of what he refers to as ``4-space,'' a $4$-dimensional analytic manifold $\mathfrak{M}$, coordinatized by $(x^{\mu})_{\mu \in \{0, 1, 2, 3\}} \in \mathbb{R}^4$, less the line $\mathcal{L} := \bigl\{(x^0, x^1, x^2, x^3) \, \big| \, x^0 \in \mathbb{R}, \, x^1 = x^2 = x^3 = 0\bigr\}$, which is the timeline of the point particle residing at the origin, and, on the other hand, by an analytic metric $\boldsymbol{g}$ satisfying the following eight physical conditions:
\begin{itemize}
\item[($\mathcal{C}1$)] Solution of the vacuum Einstein field equations $R_{\mu \nu}(\boldsymbol{g}) = 0$ for all $\mu, \nu \in \{0, 1, 2, 3\}$.
\item[($\mathcal{C}2$)] Invariant under the $1$-parameter group of time translations
\begin{equation*}
T_t \colon 
\begin{cases}
\, x^0 \mapsto \overline{x}^0 = x^0 - t \\[0.25cm]
\, \hspace{0.03cm} x^i \mapsto \overline{x}^i = x^i \, ,
\end{cases}
\end{equation*}
where $i \in \{1, 2, 3\}$.
\item[($\mathcal{C}3$)] Invariant under the connected $3$-parameter group of spatial rotations
\begin{equation*}
R_r \colon 
\begin{cases}
\, x^0 \mapsto \overline{x}^0 = x^0 \\[0.25cm]
\, \hspace{0.03cm} x^i \mapsto \overline{x}^i = r_j^{\,\,\, i} \, x^j 
\end{cases}
\end{equation*}
with $\boldsymbol{r}^{\textnormal{T}} \boldsymbol{r} = \1$ and $\textnormal{det}(\boldsymbol{r}) = 1$.
\item[($\mathcal{C}4$)] Asymptotic to the Minkowski metric at spatial infinity, i.e., $g_{\mu \nu} \rightarrow \eta_{\mu \nu}$ for $x^i x_i \rightarrow \infty$.\footnote{Instead of ``Minkowski metric,'' Finkelstein calls this metric the \textit{Lorentz} metric.}
\item[($\mathcal{C}5$)] Not extendable to the line $\mathcal{L}$ (except for the trivial case $g_{\mu \nu} \equiv \eta_{\mu \nu}$).\footnote{For it is the world line of the gravitational-field-generating point particle, Finkelstein regards this line as a ``true singularity.''} 
\item[($\mathcal{C}6$)] Invariant under the discrete group of spatial reflections 
\begin{equation*}
P \colon 
\begin{cases}
\, x^0 \mapsto \overline{x}^0 = x^0 \\[0.25cm]
\, \hspace{0.03cm} x^i \mapsto \overline{x}^i = - x^i \, .
\end{cases}
\end{equation*}
\item[($\mathcal{C}7$)] Invariant under the discrete group of time reversals
\begin{equation*}
T \colon
\begin{cases}
\, x^0 \mapsto \overline{x}^0 = - x^0 \\[0.25cm]
\, \hspace{0.03cm} x^i \mapsto \overline{x}^i = x^i \, .
\end{cases}
\end{equation*}
\item[($\mathcal{C}8$)] The component $g_{0 0} > 0$ throughout $\mathfrak{M}$ for $x^0$ to be timelike.
\end{itemize}
He continues with proving that for a spherically symmetric point particle of \textit{positive} mass, one already obtains a unique analytic solution 
\begin{equation} \label{FSM} 
\boldsymbol{g} = \biggl(1 - \frac{1}{r}\biggr) \textnormal{d}x^0 \otimes \textnormal{d}x^0 - \biggl(1 + \frac{1}{r}\biggr) \textnormal{d}r \otimes \textnormal{d}r + \frac{1}{r} \, [\textnormal{d}x^0 \otimes \textnormal{d}r + \textnormal{d}r \otimes \textnormal{d}x^0] - r^2 \boldsymbol{g}_{S^2}    
\end{equation}
for coordinates $(x^0, r, \vartheta, \phi) \in \mathbb{R} \times (1, \infty) \times (0, \pi) \times [0, 2 \pi)$ by imposing only Conditions ($\mathcal{C}1$)--($\mathcal{C}5$).\footnote{Actually, Finkelstein presents an equivalent mixed Cartesian-spherical polar coordinate representation of this solution.}$^{,}$\footnote{The Cartesian coordinates $(x^1, x^2, x^3) \in \mathbb{R}^3 \backslash \{\boldsymbol{0}\}$ on which Finkelstein's spherical polar coordinates are based are all normalized to $2 M$, resulting in a ``mass-free'' representation of the metric.}$^{,}$\footnote{For the proof, Finkelstein defines the metric directly and then verifies the validity of Conditions ($\mathcal{C}1$)--($\mathcal{C}5$). The uniqueness of a metric satisfying these conditions is taken as a well-known result. He does not, however, use the conditions to determine the metric in a constructive manner as claimed.} This metric automatically satisfies Condition ($\mathcal{C}6$), but it is incompatible with Conditions ($\mathcal{C}7$) and ($\mathcal{C}8$).\footnote{For the details of Finkelstein's reasoning see \cite{Finkelstein}.} Particularly, Finkelstein's famous coordinate transformation for the exterior Schwarzschild solution enters the step in his proof where he establishes the validity of Condition ($\mathcal{C}1$). More precisely, Finkelstein first maps the metric (\ref{FSM}) into the usual Schwarzschild representation employing the coordinate transformation 
\begin{equation} \label{FinkelsteinTrafo} 
\mathfrak{T}^{(\textnormal{F})} \colon
\begin{cases}
\, \mathbb{R} \times (1, \infty) \times (0, \pi) \times [0, 2 \pi) \rightarrow \mathbb{R} \times (1, \infty) \times (0, \pi) \times [0, 2 \pi) \\[0.25cm]
\hspace{2.67cm} (x^0, r, \vartheta, \phi) \mapsto (\bar{x}^0, \bar{r}, \vartheta', \phi') 
\end{cases} 
\end{equation}
with
\begin{equation*}
\bar{x}^0 = x^0 + \ln{(r - 1)} \, , \quad \bar{r} = r \, , \quad \vartheta' = \vartheta \, , \quad \textnormal{and} \quad \phi' = \phi \, .\footnote{Finkelstein formulates the spatial part of this coordinate transformation in terms of Cartesian coordinates rather than in spherical polar coordinates.}^{,}\footnote{This coordinate transformation gives rise to the \textit{normalized} Schwarzschild representation
\begin{equation*} 
\boldsymbol{g} = \biggl(1 - \frac{1}{\bar{r}}\biggr) \textnormal{d}\bar{x}^0 \otimes \textnormal{d}\bar{x}^0 - \biggl(1 - \frac{1}{\bar{r}}\biggr)^{- 1} \textnormal{d}\bar{r} \otimes \textnormal{d}\bar{r} - \bar{r}^2 \boldsymbol{g}_{S^2} 
\end{equation*}
of the Schwarzschild metric.}
\end{equation*}
Then, since the Schwarzschild representation of the exterior Schwarzschild solution satisfies the vacuum Einstein field equations $\overline{R}_{\mu \nu} = 0$ for $\overline{r} > 1$ and all $\mu, \nu \in \{0, 1, 2, 3\}$, he infers that the same holds also in the Finkelstein representation, i.e., $R_{\mu \nu} = 0$ for $r > 1$, as $\overline{R}_{\mu \nu}$ is a tensor. And because $R_{\mu \nu}$ is analytic in $\mathfrak{M}$, he finally concludes that $R_{\mu \nu} = 0$ everywhere in $\mathfrak{M}$.
\begin{Remark}
Since Finkelstein coordinates are, except for a normalization, identical to Eddington coordinates, they also have the properties described in Remark \ref{EddProps}. However, Finkelstein explicitly considers an analytic extension of his coordinates into the (nonstatic) region with $r < 1$---one that is regular across the past component of the event horizon at $r = 1$---thereby covering the region $\textnormal{B}_{\textnormal{I}} \cup \textnormal{B}_{\textnormal{IV}}$.\footnote{Finkelstein addresses the existence of another ``distinct completion'' of the exterior Schwarzschild solution, viz., the time reversed of this analytic extension, which relates to a coordinate system that has a time coordinate $\bar{x}^0 = x^0 - \ln{|r - 1|}$ and that covers the region $\textnormal{B}_{\textnormal{I}} \cup \textnormal{B}_{\textnormal{II}}$. Therefore, he views the process of gravitational collapse of, e.g., a star, as a random transition from a manifold consisting solely of the exterior region $\textnormal{B}_{\textnormal{I}}$ to one of the two distinct but equivalent analytic extensions given by $\textnormal{B}_{\textnormal{I}} \cup \textnormal{B}_{\textnormal{II}}$ and $\textnormal{B}_{\textnormal{I}} \cup \textnormal{B}_{\textnormal{IV}}$. He furthermore remarks that during the collapse, a physical singularity forms only at the origin at $r = 0$ [the Schwarzschild singularity located at $r = 1$ is removed through (the inverse of) Finkelstein's coordinate transformation (\ref{FinkelsteinTrafo})], whereas the hypersurface at $r = 1$ is now identified as a perfect unidirectional membrane, i.e., a surface that can be traversed by causal effects in only one direction, with the respective unidirectionalities of the membranes of the two distinct analytic extensions reversed.}
\end{Remark}
\begin{Remark}
Finkelstein argues for a past-future asymmetry of the gravitational field of a massive, spherically symmetric point particle, which he essentially attributes to an ``instability'' caused by the nonlinearity of the theory, on the basis of the corresponding spacetime, although being analytic and complete, featuring a metric that is not time-reversal invariant [cf.\ Equation (\ref{FSM})]. Nonetheless, at the end of \cite{Finkelstein}, he included a late footnote containing a ``Note added in proof'' that contradicts this view:
\begin{quote}
Schild points out that $\mathfrak{M}$ is still incomplete, it possesses a nonterminating geodesic of finite length in one direction. Kruskal has sketched for me a manifold $\mathfrak{M}^{\star}$ that is complete and contains $\mathfrak{M}$. $\mathfrak{M}^{\star}$ is time-symmetric and violates one of the conditions on $\mathfrak{M}$: it does not have the topological structure of all of $4$-space less a line. Kruskal obtained $\mathfrak{M}^{\star}$ some years ago (unpublished).     
\end{quote}
This realization had to ultimately foil his conviction that for a massive, spherically symmetric point particle the concept of time invariance and the theory of general relativity are incompatible, for Kruskal's result explicitly showed that the argumentation for a past-future asymmetry is grounded on the erroneous premise of the completeness of the underlying manifold $\mathfrak{M}$, rendering the supposed asymmetry a mere artifact ascribed to a nonexhaustive choice of manifold. Kruskal's clarifying paper \cite{Kruskal} on the matter was eventually published in 1960 (see Section \ref{SubsectionIIB}).
\end{Remark}

\subsubsection{Penrose Coordinates} \label{SubsubsectionVC}

\noindent In his 1969 paper \cite{Penrose3} on the process of gravitational collapse of a star in general relativity, which is in a sense a popularization of his famous 1965 singularity theorem paper \cite{Penrose2}, Penrose constructs an analytic extension of the exterior Schwarzschild spacetime across the future component of its event horizon as a simple vacuum model of the interior region of the contracting star.\footnote{An adequate description of the star's interior is required for a \textit{complete} account of the process of gravitational collapse.} For this purpose, he utilizes the transformation from Schwarzschild coordinates into horizon-penetrating coordinates characterized by what he calls an advanced time parameter
\begin{equation*} 
\mathfrak{T}^{(\textnormal{P})} \colon
\begin{cases}
\, \mathbb{R} \times (2 M, \infty) \times (0, \pi) \times [0, 2 \pi) \rightarrow \mathbb{R} \times (2 M, \infty) \times (0, \pi) \times [0, 2 \pi)  \\[0.25cm]
\hspace{3.25cm} (t, r, \vartheta, \phi) \mapsto (v, r', \vartheta', \phi') 
\end{cases} 
\end{equation*}
with 
\begin{equation} \label{PCT} 
v = t + r + 2 M \ln{(r - 2 M)} \, , \quad r' = r \, , \quad \vartheta' = \vartheta \, , \quad \textnormal{and} \quad \phi' = \phi \, .
\end{equation}
In terms of these coordinates, the exterior Schwarzschild metric takes the form
\begin{equation*}
\boldsymbol{g} = \biggl(1 - \frac{2 M}{r}\biggr) \textnormal{d}v \otimes \textnormal{d}v - \textnormal{d}v \otimes \textnormal{d}r - \textnormal{d}r \otimes \textnormal{d}v - r^2 \, \boldsymbol{g}_{S^2} \, ,
\end{equation*}
which can be analytically extended from the exterior region $\textnormal{B}_{\textnormal{I}}$ across the future component of the event horizon at $r = 2 M$ into the (nonstatic) interior region $\textnormal{B}_{\textnormal{II}}$, thus making it regular for all $r \in \mathbb{R}_{> 0}$.\footnote{Even though Penrose did not work out the full coordinate transformation for the analytic extension across the future component of the event horizon, where the advanced time parameter $v$ in the extending region has to be of the particular form $v = - t + r + 2 M \ln{(2 M - r)}$, he nevertheless points out that the Schwarzschild metric in this representation accounts for both regions $\textnormal{B}_{\textnormal{I}}$ and $\textnormal{B}_{\textnormal{II}}$.}
\begin{Remark} \label{RemarkPenrose}
Unlike Eddington--Finkelstein coordinates, which feature, besides three spatial coordinates, a time coordinate that is a proper temporal function, Penrose's coordinates, although similar in appearance (except for the different sign in front of the logarithmic contribution and the additional linear term $+ r$ both in the advanced time parameter $v$), are indeed null coordinates, i.e., in addition to the same three spatial coordinates as in the Eddington--Finkelstein case, they comprise a null coordinate, $v$, instead of a time coordinate. The fact that Penrose's advanced time parameter is in reality a null coordinate can be traced back to it being adapted to the tangent vectors of the ingoing radial null geodesics in the exterior Schwarzschild spacetime expressed in Schwarzschild coordinates
\begin{equation*}
\frac{\textnormal{d}t}{\textnormal{d}s} = \frac{r^2}{\Delta} \, \mathcal{E}_0 \quad \textnormal{and} \quad \frac{\textnormal{d}r}{\textnormal{d}s} = - \mathcal{E}_0 \, , 
\end{equation*}
where $s$ is an affine parameter and $\mathcal{E}_0 \in \mathbb{R}_{> 0}$ a constant. More precisely, eliminating the affine parameter gives rise to the relation 
\begin{equation*}
\frac{\textnormal{d}t}{\textnormal{d}r} = - \biggl(1 - \frac{2 M}{r}\biggr)^{- 1} \quad \Leftarrow \quad t = - r - 2 M \ln(r - 2 M) + \mathcal{C}_0   
\end{equation*}
between the Schwarzschild time and radial coordinates, with $\mathcal{C}_0 \in \mathbb{R}$ being a constant of integration, which may be viewed as the basis for the definition of the coordinate $v$ specified in Equation (\ref{PCT}).
\end{Remark}

\subsection{Comparison of the Eddington--Finkelstein and Penrose Foliations} \label{SubsectionVB}

\noindent Now, to demonstrate the usefulness of Penrose diagrams, we bring out the differences between the Eddington--Finkelstein and Penrose coordinate representations of the Schwarzschild spacetime, which we have already briefly addressed in Remark \ref{RemarkPenrose}, by visually analyzing---and then contrasting---the Penrose diagram of the maximally analytically extended Schwarzschild spacetime with the region $\textnormal{B}_{\textnormal{I}} \cup \textnormal{B}_{\textnormal{II}}$ foliated by the level sets of the time-reversed (or advanced) Eddington--Finkelstein time coordinate $v_{\textnormal{EF}} = t + 2 M \ln|r/(2 M) - 1|$, on the one hand, and by the level sets of the Penrose null coordinate $v_{\textnormal{Pen}} = v_{\textnormal{EF}} + r$, on the other hand.\footnote{For consistency with Section \ref{SubsectionIVA}, where the Regge--Wheeler coordinate and derived quantities have consistent physical units, we here employ variants of the Eddington--Finkelstein time and Penrose null coordinates that comprise the extra additive constant $- 2 M \ln{(2 M)}$.} This approach has the advantage that the differences can be directly read off without resorting to any mathematical formalisms.

We begin with the analysis of the Penrose diagram in FIG.\ 2(a) depicting a representative subset of the family of level sets of the Eddington–Finkelstein time coordinate $v_{\textnormal{EF}}$. These level sets are all spacelike (at every point they have an opening angle with the abscissa smaller than $45^{\circ}$) and extend from spacelike infinity $i^0$ to the curvature singularity at $r = 0$, where they terminate. They cross smoothly from the exterior region $\textnormal{B}_{\textnormal{I}}$ through the future component of the event horizon at $r = 2 M$ into the black hole region $\textnormal{B}_{\textnormal{II}}$, always retaining their spacelike character. In contrast, a direct visual inspection of the Penrose diagram in FIG.\ 2(b) that contains a representative subset of the family of level sets of the Penrose null coordinate $v_{\textnormal{Pen}}$ shows these hypersurfaces to be null (with $45^{\circ}$ opening angles at every point), extending smoothly and causal-type-preserving from past null infinity $\mathscr{I}^-$ through the future component of the event horizon up to their termination points at the curvature singularity. The Penrose coordinates defined in \cite{Penrose3} are therefore the time-reversed null variant of the Eddington--Finkelstein coordinates introduced in \cite{Eddington, Finkelstein}.

\begin{figure}[t]%
\centering
\subfigure[][]{%
\label{PenroseKruskalEF}%
\includegraphics[width=0.48\columnwidth]{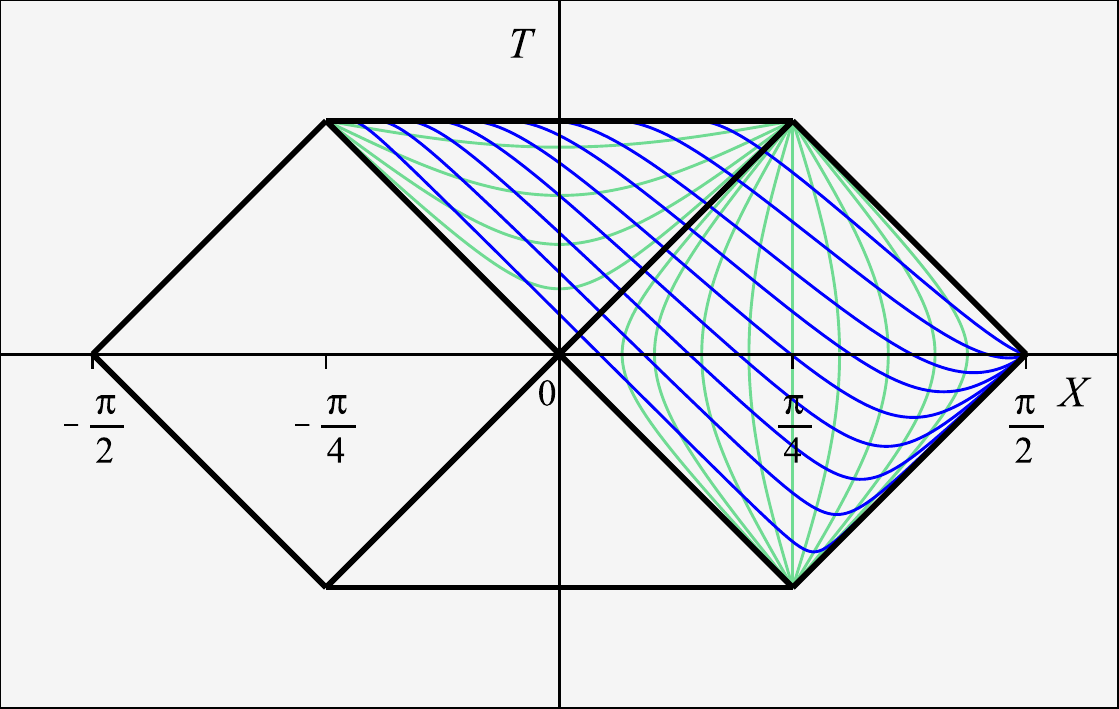}}
\hspace{8pt}%
\subfigure[][]{%
\label{PenroseKruskalPen}%
\includegraphics[width=0.48\columnwidth]{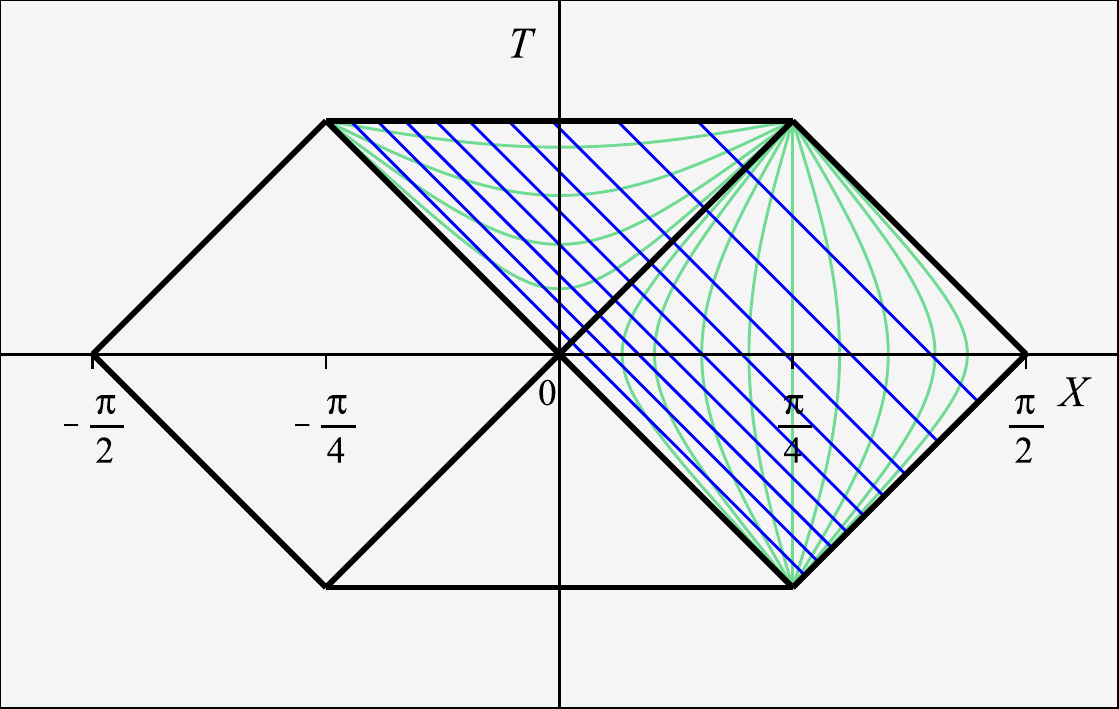}} \\
\subfigure[][]{%
\label{PenroseKruskalS2}%
\includegraphics[width=0.48\columnwidth]{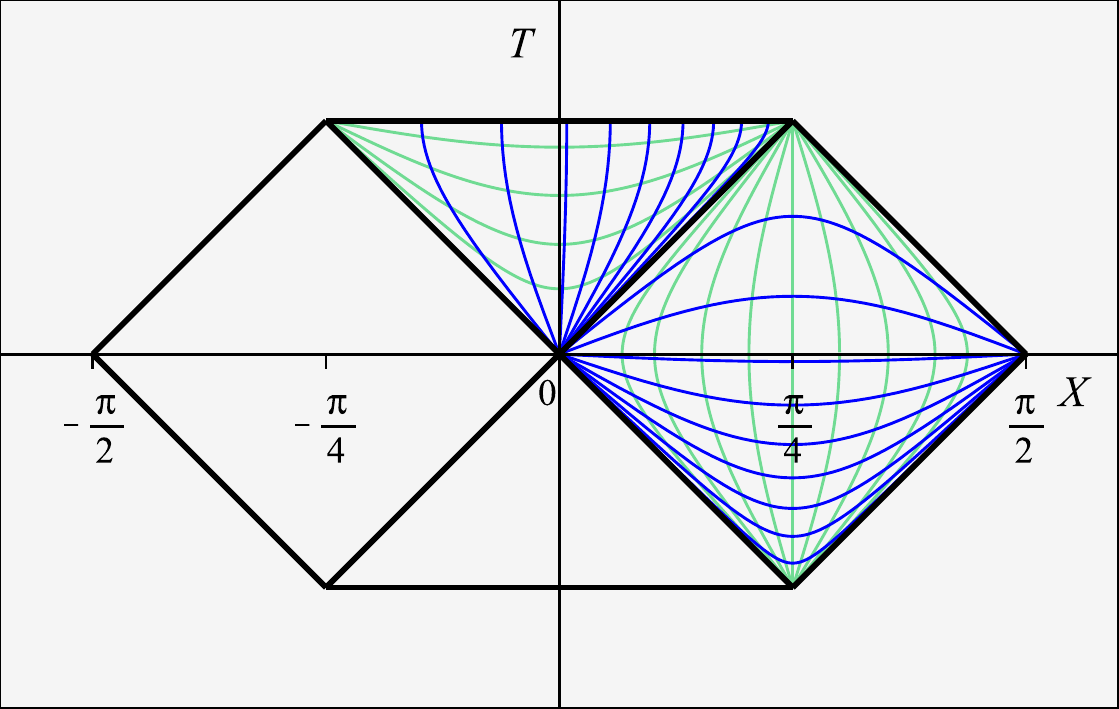}}
\hspace{8pt}%
\subfigure[][]{%
\label{PenroseKruskalKS}%
\includegraphics[width=0.48\columnwidth]{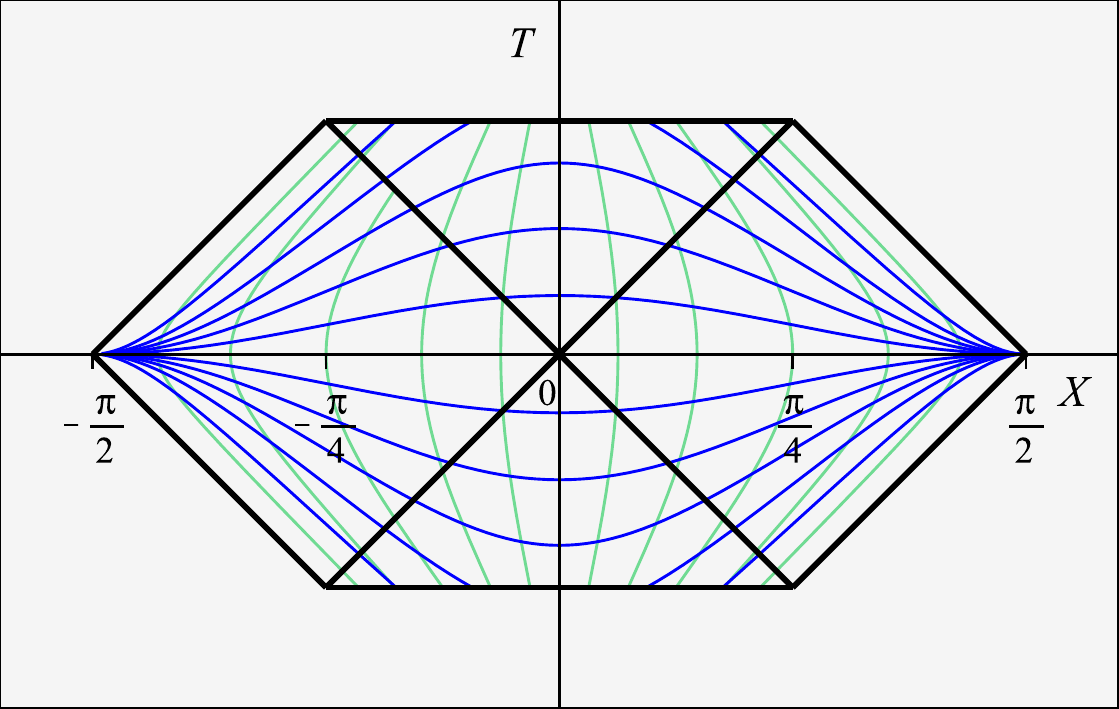}}
\caption[...]
{\subref{PenroseKruskalEF} Penrose diagram of the maximally analytically extended Schwarzschild spacetime with level sets of the Eddington--Finkelstein time coordinate $v_{\textnormal{EF}}$ (blue curves) and level sets of the Eddington--Finkelstein radial coordinate $r$ (aquamarine curves). \subref{PenroseKruskalPen}--\subref{PenroseKruskalKS} The same diagram for the Penrose representation, the Schwarzschild representation extended to the region $\textnormal{B}_{\textnormal{I}} \cup \textnormal{B}_{\textnormal{II}}$, and the Kruskal representation, respectively.}%
\label{PenroseKruskalSlice}%
\end{figure}

%
%
\begin{Remark}
For the construction of the family of level sets of, e.g., the Eddington--Finkelstein time coordinate $v_{\textnormal{EF}}$ in the region $\textnormal{B}_{\textnormal{I}} \cup \textnormal{B}_{\textnormal{II}}$, one essentially derives a transcendental algebraic equation that relates the compactified Kruskal--Szekeres time and radial coordinates $T$ and $X$, which span the Penrose diagram, to the Eddington--Finkelstein time coordinate $v_{\textnormal{EF}}$. To this end, one inserts both the transformation laws $U = T - X$ and $V = T + X$ for the compactified Kruskal--Szekeres double-null coordinates $U$ and $V$ [cf.\ Equation (\ref{TXUV})] and the transformation laws for the Eddington--Finkelstein double-null coordinates $u$ and $v$ in the regions $\textnormal{B}_{\textnormal{I}}$ and $\textnormal{B}_{\textnormal{II}}$ specified in Equation (\ref{DNC1}) into Equation (\ref{DNC2}). Subsequently, using the transformation laws $t = v_{\textnormal{EF}} - 2 M \ln|r'/(2 M) - 1|$ and $r = r'$ between the Schwarzschild time and radial coordinates $t$ and $r$ and the Eddington--Finkelstein time and radial coordinates $v_{\textnormal{EF}}$ and $r'$ in the resulting expressions, and applying simple algebraic manipulations, one obtains
\begin{equation*}
\left\{\!\begin{aligned}
& \, \displaystyle \tan(T \pm X) = e^{(v_{\textnormal{EF}} + r')/(4 M)} \\[0.15cm] 
& \, \displaystyle \tan(T \mp X) = e^{- (v_{\textnormal{EF}} - r')/(4 M)} \bigl[1 - r'/(2 M)\bigr]
\end{aligned}\right\} \,\,\,\, \textnormal{for} \,\,\, \textnormal{B}_{\textnormal{I}} \slash \textnormal{B}_{\textnormal{II}} \, .
\end{equation*}
Eliminating the Eddington--Finkelstein radial coordinate $r'$ immediately yields the desired relation
\begin{equation*}
2 \ln\bigl(\tan(T \pm X)\bigr) + e^{v_{\textnormal{EF}}/(2 M)} \tan(T \mp X) \cot(T \pm X) = \frac{v_{\textnormal{EF}}}{2 M} + 1 \, \hspace{0.3cm} \textnormal{for} \,\, \textnormal{B}_{\textnormal{I}} \slash \textnormal{B}_{\textnormal{II}} \, .
\end{equation*}
The family of level sets of the Eddington--Finkelstein time coordinate $v_{\textnormal{EF}}$ in the region $\textnormal{B}_{\textnormal{I}} \cup \textnormal{B}_{\textnormal{II}}$ is then given by the combined solutions $\bigl\{ T_{v_0}(X) \, \big| \, v_{\textnormal{EF}} = v_0 \,\,\, \textnormal{for all} \,\,\, v_0 \in \mathbb{R} \,\,\, \textnormal{and} \,\,\, X \in (- \pi/4, \pi/2)\bigr\}$
of these transcendental algebraic equations, where each constant $v_0$ uniquely determines one particular level set.
\end{Remark}

For completeness, in FIG.\ 2(c) and FIG.\ 2(d), we also depict the Penrose diagram of the maximally analytically extended Schwarzschild spacetime with the region $\textnormal{B}_{\textnormal{I}} \cup \textnormal{B}_{\textnormal{II}}$ foliated by the families of level sets of the Schwarzschild time and radial coordinates $(t, r) \in \mathbb{R} \times \mathbb{R}_{> 0}$ and with the entire maximal analytic extension foliated by the families of level sets of the Kruskal time and radial coordinates $(v, u) \in \mathbb{R} \times \mathbb{R}$, respectively (for the definitions of these coordinates see Section \ref{SubsectionIIB}). In the Schwarzschild case, both families of level sets are fully contained either in the exterior region $\textnormal{B}_{\textnormal{I}}$ or in the black hole region $\textnormal{B}_{\textnormal{II}}$, never traversing---but asymptotically approaching---the future component of the event horizon at $r = 2 M$.\footnote{These families can also be constructed separately in the parallel exterior region $\textnormal{B}_{\textnormal{III}}$ and the white hole region $\textnormal{B}_{\textnormal{IV}}$, for Schwarzschild coordinates can be used to cover these regions as well (the same holds for the Eddington--Finkelstein and Penrose foliations in the region $\textnormal{B}_{\textnormal{III}} \cup \textnormal{B}_{\textnormal{IV}}$).}$^{,}$\footnote{In $\textnormal{B}_{\textnormal{I}}$, the future component of the event horizon coincides with the level set of the Schwarzschild time $t = + \infty$, which reflects the fact that an infalling observer described in Schwarzschild coordinates requires an infinite amount of coordinate time to reach the horizon. For Eddington--Finkelstein and Penrose coordinates, this time interval is finite.} Their causal types, however, reverse across the event horizon, i.e., the spacelike, parabola-shaped level sets of the Schwarzschild time coordinate $t$ in $\textnormal{B}_{\textnormal{I}}$ [blue curves extending from spacelike infinity $i^0$ to the bifurcation $2$-sphere at $(T = 0, X = 0)$, which at every point have an opening angle with the abscissa smaller than $45^{\circ}$] become timelike, hyperbola-shaped level sets in $\textnormal{B}_{\textnormal{II}}$ (blue curves extending from the bifurcation $2$-sphere to the curvature singularity at $r = 0$ with opening angles larger than $45^{\circ}$ at every point), and vice versa for the level sets of the Schwarzschild radial coordinate $r$ [aquamarine curves hyperbolically extending from past timelike infinity $i^-$ to future timelike infinity $i^+$ in $\textnormal{B}_{\textnormal{I}}$ and parabolically extending from $(t = - \infty, r = 0)$ to $(t = + \infty, r = 0)$ in $\textnormal{B}_{\textnormal{II}}$]. In the Kruskal case, the main part of the spacelike level sets of the Kruskal time coordinate $v$ (blue curves with opening angles smaller than $45^{\circ}$ at every point) begins at spatial infinity $i^0$ in $\textnormal{B}_{\textnormal{I}}$, crosses into---and traverses smoothly---either $\textnormal{B}_{\textnormal{II}}$ or $\textnormal{B}_{\textnormal{IV}}$, and ends at spatial infinity $i^0$ in $\textnormal{B}_{\textnormal{III}}$. But there also exist disjoint, mirror-symmetrical pairs of spacelike level sets of the Kruskal time coordinate with components that originate at $i^0$ in $\textnormal{B}_{\textnormal{I}}$ and $\textnormal{B}_{\textnormal{III}}$, respectively, and terminate at the curvature singularity at $\bigl\{(T, X) \, \big| \, T = \pm \pi/4 \,\,\, \textnormal{and} \,\, - \pi/4 < X < \pi/4\bigr\}$. These particular level sets occur when the Kruskal time coordinate either exceeds the upper critical value $v = 1$ or is below the lower critical value $v = - 1$, and can be explained by the constraint $v^2 < 1 + u^2$, which applies in $\textnormal{B}_{\textnormal{II}}$ and $\textnormal{B}_{\textnormal{IV}}$ (cf.\ the end of Section \ref{SubsectionIIB}), being violated for specific ranges of the values of $u$. Both kinds of level sets retain their causal type throughout the entire analytic extension. The timelike level sets of the Kruskal radial coordinate $u$ (aquamarine curves with opening angles larger than $45^{\circ}$ at every point), on the other hand, extend smoothly from the past curvature singularity at $\bigl\{(T, X) \, \big| \, T = - \pi/4 \,\,\, \textnormal{and} \,\, - \pi/4 < X < \pi/4\bigr\}$ in $\textnormal{B}_{\textnormal{IV}}$, passing either through $\textnormal{B}_{\textnormal{I}}$ or $\textnormal{B}_{\textnormal{III}}$, to the future curvature singularity at $\bigl\{(T, X) \, \big| \, T = \pi/4 \,\,\, \textnormal{and} \,\, - \pi/4 < X < \pi/4\bigr\}$ in $\textnormal{B}_{\textnormal{II}}$, also always preserving their causal types.

\vspace{0.2cm}

\section*{Acknowledgments}

\noindent The author is grateful to Dennis Lehmkuhl for suggesting to write a pedagogical paper on the construction and use of Penrose diagrams, and to work out the differences between the Eddington--Finkelstein and Penrose coordinate representations of the Schwarzschild spacetime. He is also grateful to the Lichtenberg Group for History and Philosophy of Physics of the University of Bonn for useful discussions on the topic, and for detailed comments on the first draft of this paper. This project has received funding from the European Research Council (ERC) under the European Union's \textit{Horizon Europe} research and innovation program (COGY, grant agreement $\textnormal{n}^{\circ} 101088528$) and from the Lichtenberg Grant for History and Philosophy of Physics of the Volkswagen Foundation.

\newpage


\end{document}